\documentclass[twocolumn]{aastex62}
\usepackage[utf8]{inputenc}
\usepackage{amsmath}
\usepackage{amssymb}
\usepackage[version=4]{mhchem}
\usepackage{xfrac}
\usepackage{graphicx}
\usepackage{ulem}

\graphicspath{{./}}

\shorttitle{Oceans \& the Sulfur Cycle}

\shortauthors{Loftus, Wordsworth, \& Morley}

\begin{document}
\title{Sulfate Aerosol Hazes and \ce{SO2} Gas as Constraints on Rocky Exoplanets' Surface Liquid Water}

\correspondingauthor{Kaitlyn Loftus}\email{kloftus@g.harvard.edu}

\author{Kaitlyn Loftus}
\affil{Department of Earth and Planetary Sciences, Harvard University, Cambridge, MA 02140, USA}

\author{Robin D. Wordsworth}
\affil{Department of Earth and Planetary Sciences, Harvard University, Cambridge, MA 02140, USA}
\affil{Harvard Paulson School of Engineering and Applied Sciences, Harvard University, Cambridge, MA 02140, USA}

\author{Caroline V. Morley}
\affil{Department of Astronomy, University of Texas at Austin, Austin, TX 78712, USA}

\begin{abstract}
Despite surface liquid water's importance to habitability, observationally diagnosing its presence or absence on exoplanets is still an open problem. Inspired within the Solar System by the differing sulfur cycles on Venus and Earth, we investigate thick sulfate (\ce{H2SO4-H2O}) aerosol haze and high trace mixing ratios of \ce{SO2} gas as observable atmospheric features whose sustained existence is linked to the near-absence of surface liquid water. 
We examine the fundamentals of the sulfur cycle on a rocky planet with an ocean and an atmosphere in which the dominant forms of sulfur are \ce{SO2} gas and \ce{H2SO4-H2O} aerosols (as on Earth and Venus). We build a simple but robust model of the wet, oxidized sulfur cycle to determine the critical amounts of sulfur in the atmosphere-ocean system required for detectable levels of \ce{SO2} and a detectable haze layer. We demonstrate that for physically realistic ocean pH values (pH $\gtrsim$ 6) and conservative assumptions on volcanic outgassing, chemistry, and aerosol microphysics, surface liquid water reservoirs with greater than $10^{-3}$ Earth oceans are incompatible with a sustained observable \ce{H2SO4-H2O} haze layer and sustained observable levels of \ce{SO2}. Thus, we propose the observational detection of an \ce{H2SO4-H2O} haze layer and of \ce{SO2} gas as two new remote indicators that a planet does not host significant surface liquid water. 
\end{abstract}

\keywords{exoplanet atmospheres --- exoplanet surface characteristics --- habitable planets}

\section{Introduction}\label{sec:intro}

Surface liquid water is considered essential to Earth-like life \citep[e.g.,][]{Scalo2007,Kasting2012}, so determining whether a planet possesses liquid water is essential to constraining its habitability. As exoplanet detection and characterization techniques improve, observational constraints on the presence of surface water in the form of oceans will be a precursor to building an understanding of the occurrence rate of Earth-like planets and, ultimately, to building an understanding of the occurrence rate of Earth-like life. For many planets, the absence of surface oceans can be determined from straightforward physical arguments. Oceans require a surface temperature and pressure consistent with the stability of liquid water. These requirements limit possible ocean-hosting planet candidates to low-mass ($M\lesssim 10 M_\oplus$) planets in the habitable zone \citep[e.g.,][]{Kasting1993,Kopparapu2013}.

While these fundamental requirements can be reasonably evaluated from a planet's density and received stellar flux, evaluating whether rocky, habitable-zone planets actually possess surface oceans will be a substantial challenge, even for next-generation telescopes. While atmospheric spectra can identify water vapor in a planet's atmosphere \citep[e.g.,][]{Deming2013,Huitson2013,Fraine2014,Sing2016}, such a detection says nothing conclusively about surface liquid water. Temporally resolved reflected light spectra can potentially identify the presence of liquid water via color variation due to dark ocean color \citep{Cowan2009}, polarized light variation due to ocean smoothness \citep{Zugger2010}, and/or ocean glint due to ocean reflectivity \citep{Robinson2010}. However, these methods all involve some combination of caveats, including still-unresolved potential false positives, decades-away instrumentation, and a reliance on ideal planetary conditions \citep{Cowan2012}. To systematically probe the presence of surface oceans on exoplanets, additional methods will unquestionably be needed. Here, we propose two such methods.

Though all existing ocean detection proposals exploit liquid water's radiative properties, other planetary-scale implications of the presence of surface liquid water exist. A hydrosphere can substantially alter a planet's chemistry, including---most notably here---its sulfur cycle. Previous studies of the observability of sulfur in exoplanet atmospheres suggest that sulfur has the capability to be observed in atmospheres in both aerosol and gas species \citep{Kaltenegger2010a,Kaltenegger2010b,Hu2013,Misra2015,Lincowski2018}. In this paper, we test whether the observations of a permanent sulfate (\ce{H2SO4-H2O}) haze layer and atmospheric \ce{SO2} can diagnose the absence of significant surface liquid water. (Note that we do not consider the inverse of this hypothesis---i.e., a lack of observable atmospheric sulfur implies the presence of an ocean---which is not particularly tenable.)

\ce{H2SO4-H2O} aerosols with sufficient optical depth could be detected in exoplanet atmospheric spectra in the near future \citep{Hu2013,Misra2015}. Within the Solar System, Venus and Earth are examples of rocky planets with drastically different surface liquid water volumes and sulfur aerosol opacities. 
Venus has a global, optically thick \ce{H2SO4-H2O} aerosol layer \citep{Knollenberg1980} and no surface liquid water. 
Earth, in contrast, only briefly hosts \ce{H2SO4-H2O} aerosol layers of significant optical depth after large volcanic eruptions \citep{Mccormick1995}. Earth's present inability to sustain an \ce{H2SO4-H2O} aerosol layer in its atmosphere can be directly tied to the presence of oceans, as we show in Section \ref{sec:methods}. 

Beyond aerosols, \ce{SO2} gas also has the potential to be detected at $\gtrsim$ 1 ppm mixing ratios \citep{Kaltenegger2010a,Hu2013}. Previous results have found that building up sulfur concentrations to this level with an Earth-like sulfur cycle requires implausibly high outgassing rates \citep{Kaltenegger2010b,Hu2013}. We show in Section \ref{sec:methods} that the presence of surface liquid water represents a fundamental barrier to maintaining high trace atmospheric sulfur gas levels in such circumstances. 

The structure of this paper is as follows. In Section \ref{sec:def}, we define the range of planetary atmospheres we consider. Section \ref{sec:methods} describes our model of the sulfur cycle on a wet, oxidized world. Section \ref{sec:results} presents the key findings of our model on the incompatibility of observable \ce{H2SO4-H2O} aerosols and \ce{SO2} gas with abundant surface liquid water. 
In Section \ref{sec:dis}, we discuss the implications and limitations of this study and make some suggestions for future work. Section \ref{sec:conclusion} summarizes our conclusions.

\begin{figure*}
\centering
\plotone{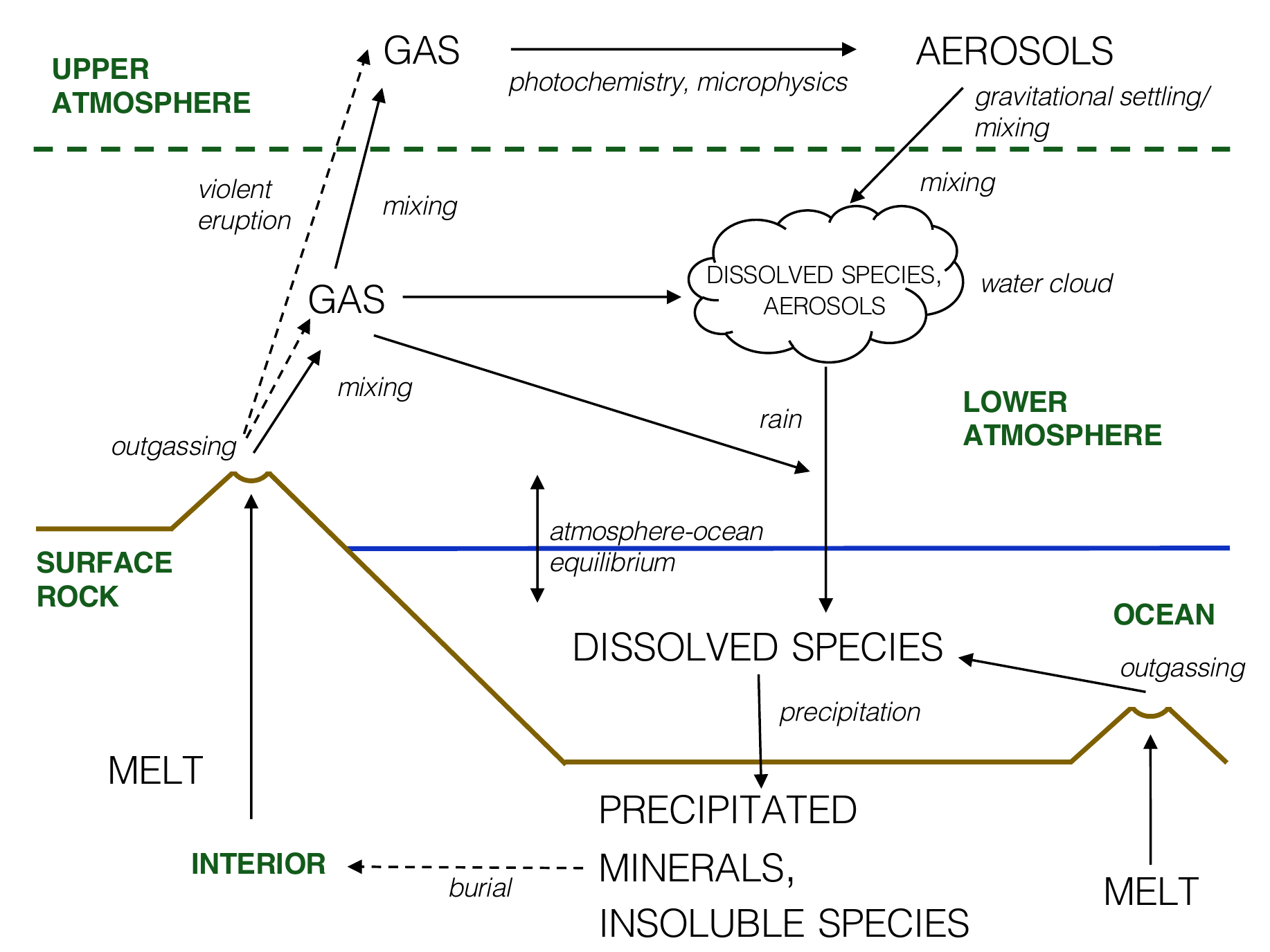}
\caption{Schematic of the major components of the sulfur cycle on a planet with an ocean and active hydrological cycle.}
\label{fig:wet_sulfur_cycle}
\end{figure*}
\section{The Sulfur Cycle on Wet, Oxidized Planets} \label{sec:def} 
In this paper, we model the sulfur cycle on wet, oxidized planets. 
We take a wet planet to be a planet with a reservoir of surface liquid water and an active hydrological cycle. We define an oxidized planet based on the atmospheric gas and aerosol sulfur species present, as discussed in further detail below.

Figure \ref{fig:wet_sulfur_cycle} summarizes the sulfur cycle on a wet planet. Sulfur gas (\ce{SO2} or \ce{H2S}) is outgassed from melt transported from the interior via volcanism. In the ocean, this gas dissolves, and the aqueous sulfur species are dynamically mixed; in the lower atmosphere, the gas is dynamically mixed.
Some gas reaches the upper atmosphere where a series of photochemical reactions produces sulfur gas that condenses into aerosols (\ce{H2SO4}-\ce{H2O} or \ce{S8}, depending on the atmospheric redox state). These aerosols are transported from the upper atmosphere to the lower atmosphere via gravitational settling or mixing on timescales that may depend on the size of the aerosol. Both sulfur gas and aerosols are removed from the lower atmosphere via their interactions with cloud-water and rainwater. Henry's law dictates an equilibrium between sulfur gas in the atmosphere and aqueous dissolved sulfur species. 
Further reactions lead to the formation of sulfur minerals, which precipitate out of the ocean when they become saturated. These sediments may eventually be recycled in the interior depending on the planet's tectonic regime.

Because sulfur can stably occupy a wide range of redox states (commonly -2, 0, +4, +6; or S(-II), S(0), S(IV), S(VI), respectively), the precise sulfur species involved in a planet's cycle depend on the oxidization state of its interior, atmosphere, and ocean \citep[e.g.,][]{Kasting1989,Pavlov2002,Johnston2011,Hu2013}. Generally, oxidized atmospheres have \ce{SO2} gas and form aerosols composed of \ce{H2SO4} and \ce{H2O}, whereas reduced atmospheres have \ce{H2S} gas and aerosols made of \ce{S8} \citep{Hu2013}. A planetary system in a more intermediate redox state will have both types of sulfur species present, with ratios between them determined by the precise oxidation level \citep{Hu2013}.

In this work, we focus on the oxidized sulfur cycle, as (1) it is much more tightly coupled to planetary water inventory and (2) its aerosol photochemistry is better constrained by observations on Earth. We consider an oxidized
atmosphere to be one in which most sulfur aerosols present are composed of \ce{H2SO4}-\ce{H2O} rather than \ce{S8}. The issue of how oxidized and reduced sulfur cycles can be distinguished observationally via transit spectra is discussed in Section \ref{subsec:oxidation_state}.

\section{Methods}\label{sec:methods}
As implied by Figure \ref{fig:wet_sulfur_cycle}, the sulfur cycle involves atmospheric dynamics, photochemistry, microphysics, aqueous chemistry, and interior dynamics. Here, we aim to combine these components into a model that maximizes simplicity while retaining all of the most essential processes. Our emphasis on simplicity is driven by our focus on the implications of the sulfur cycle for the detection of an ocean. Furthermore, as we discuss throughout this section, the large uncertainties associated with many of these processes mean that more complicated models would be unlikely to give significantly more accurate results. 
 
In our model, key parameters are constrained from basic physical arguments and Solar System analogs. For each parameter $x$, we consider both the best reasonable estimate ($x^{\text{b}}$) and the limiting scenario that promotes conditions for observable sulfur ($x^\ell$). The latter presents the most challenging conditions for our hypothesis that oceans are incompatible with sustained observable \ce{SO2} or \ce{H2SO4-H2O} haze and thus represents the most stringent test of our hypothesis. Table \ref{tab:parameters} summarizes the two sets of parameter values. 

Our model considers three reservoirs for surface sulfur: 
(1) an isothermal upper atmosphere (stratosphere), (2) a moist adiabatic lower atmosphere (troposphere), and (3) a surface liquid water layer (ocean). This basic atmospheric thermal structure arises from physical principles that are expected to be a reasonable approximation across a wide variety of planetary conditions \citep[e.g.,][]{Pierrehumbert2010}. 
The sulfur species and phases that we include are 
\ce{H2SO4 ($\ell$)} and \ce{SO2 (g)} in the stratosphere;
\ce{SO2 (g)} in the troposphere; and
\ce{SO2 (aq)}, \ce{HSO3-(aq)}, and \ce{SO3^2- (aq)}---S(IV) species---in the ocean. The planetary parameters required are surface pressure $p_\text{surf}$, surface temperature $T_\text{surf}$, stratosphere temperature $T_\text{strat}$, atmospheric composition, planetary radius $R_\text{P}$, planetary mass $M_\text{P}$, and surface relative humidity (saturation of water vapor) $\text{RH}_\text{surf}$. 

Our modeling approach begins with the critical \ce{SO2} mixing ratio for detection of atmospheric \ce{SO2} or the critical aerosol optical depth for observational detection of a haze layer. Working backward, we then calculate the critical number of sulfur atoms in the atmosphere and ocean system necessary for detection. Finally, we compare this critical value to the expected number of surface sulfur atoms to evaluate whether observable \ce{SO2} buildup or observable haze formation is likely. 
In the following subsections, we discuss the various components of our model in detail. 

\subsection{Critical Atmospheric Sulfur for Observable \ce{SO2}}
\ce{SO2} gas is observable because it absorbs in characteristic bands in the infrared. In order to be observed, \ce{SO2} must be present in high-enough concentrations such that its absorption lines have sufficient width and strength to be identified. \cite{Kaltenegger2010a} find a constant \ce{SO2} mixing ratio of 1-10 ppm observable via transit spectroscopy in an Earth-like atmosphere. 
We thus set the critical \ce{SO2} mixing ratio for detection $f^\ast_{\ce{SO2}}$ to $f^{\ast\text{b}}_{\ce{SO2}} = 1$ ppm in our reasonable case scenario and  conservatively to $f^{\ast\ell}_{\ce{SO2}} = 0.01$ ppm in our limiting case scenario. 
\cite{Hu2013} illustrate that photochemical loss of \ce{SO2} can be a significant limiting factor in the lifetime of \ce{SO2} in the atmosphere and thus can prevent high \ce{SO2} buildup. Here, we ignore the photochemical loss of \ce{SO2}---again to conservatively estimate the minimum sulfur required for detection. As shown in Section \ref{sec:results}, we can draw strong conclusions even from this minimum \ce{SO2} value, so there is little need to complicate this portion of our model with photochemistry. We discuss the photochemistry of \ce{SO2} in the context of aerosol formation in Section \ref{subsec:photochemistry}.
Finally, conceivably there may be exoplanet UV-photon-limited regimes where \ce{SO2} is neither significantly photodissociated nor destroyed from reactions with reactive photochemical products, in which case this maximum estimate would be valid. 

To first order, the amount of \ce{SO2} in the light path, not \ce{SO2}'s relative abundance in the atmosphere, is what characterizes \ce{SO2}'s spectroscopic influence \citep{Pierrehumbert2010}; mixing ratio is a convenient and traditional way of expressing abundance of trace gases, but its relationship to observability in transmission spectra will vary with total atmospheric pressure. To make \cite{Kaltenegger2010a}'s \ce{SO2} detection threshold more broadly applicable to non-Earth planetary conditions, we translate their critical mixing ratio of \ce{SO2} for detection in an Earth-like atmosphere to a critical mass column (mass per unit area) of \ce{SO2} $u^\ast_{\ce{SO2}}$:
\begin{equation}\label{eq:usurf_so2}
   u^\ast_{\ce{SO2}} = \frac{f^\ast_{\ce{SO2}} p_{\text{surf},\oplus}}{g_\oplus} \frac{\mu_{\ce{SO2}}}{\mu_{\text{air,}\oplus}}
\end{equation}
where $p_{\text{surf},\oplus}$ is the Earth's surface pressure, $g_\oplus$ is the Earth's surface gravity, $\mu_{\ce{SO2}}$ is the molar mass of \ce{SO2}, and $\mu_{\text{air,}\oplus}$ is the average molar mass of Earth air.
Evaluating Equation \eqref{eq:usurf_so2} yields $u^{\ast\text{b}}_{\ce{SO2}} = 2.3 \times 10^{-2}$ kg m$^{-2}$ and $u^{\ast\ell}_{\ce{SO2}} = 2.3 \times 10^{-4}$ kg m$^{-2}$. We then set the critical partial pressure of \ce{SO2} at the surface $p_{\ce{SO2},\text{surf}}^\ast$ as
\begin{equation}\label{eq:pso2surf_so2}
p_{\ce{SO2},\text{surf}}^\ast = u^\ast_{\ce{SO2}}g\frac{\mu_{\text{air}}}{\mu_{\ce{SO2}}}
\end{equation}
where $g$ is the local surface gravity and $\mu_{\text{air}}$ is the average molar mass of the planet's air. 
In Sections \ref{subsec:aero_extinct}-\ref{subsec:obs_aero}, we calculate the corresponding critical $p_{\ce{SO2},\text{surf}}^\ast$ for observable \ce{H2SO4-H2O} aerosols, before returning in Section \ref{subsec:atmoc} to the implications of this critical $p_{\ce{SO2},\text{surf}}^\ast$ for the sulfur budget of the ocean. 

\subsection{Aerosol Extinction}\label{subsec:aero_extinct}
\ce{H2SO4-H2O} aerosols are observable because they extinguish (scatter and absorb) light. This extinction of light is most effective per unit mass of particle in the Mie scattering regime, where particles are the same order-of-magnitude size as the wavelength of light being scattered. The mass extinction coefficient $\kappa_e$ characterizes how effectively light is extinguished per unit particle mass as
\begin{equation}\label{eq:kappa}
    \kappa_\text{e} = \frac{3}{4}\frac{Q_\text{e}}{r \rho_\text{aero}},
\end{equation}
where $r$ is the average particle radius, $\rho_\text{aero}$ is the average particle density, and $Q_\text{e}$ is the particle extinction efficiency \citep{Pierrehumbert2010}.  Physically, $r$ and $\rho_\text{aero}$ are determined from the microphysics of \ce{H2SO4-H2O} aerosol formation. In contrast with water cloud formation, \ce{H2SO4} and \ce{H2O} can condense directly from the gas phase to form liquid aerosol particles at physically realizable saturation levels \citep{Seinfeld2012,Maattanen2018}. 

The ratio of \ce{H2SO4} to \ce{H2O} by mass in an aerosol $w$ depends on both temperature and the ambient number densities of \ce{H2SO4} and \ce{H2O} vapor \citep{Maattanen2018}. The ratio $w$ controls both $\rho_\text{aero}$ ($\rho_\text{aero} = w\rho_{\ce{H2SO4}} + (1-w)\rho_{\ce{H2O}}$) and the index of refraction of the aerosol, which impacts $Q_\text{e}$. Theoretically, more \ce{H2SO4} gas is predicted to lead to aerosols with higher \ce{H2SO4} concentrations \citep{Maattanen2018}; however, observational estimates for $w$ from \ce{H2SO4-H2O} aerosols in Earth's stratosphere (a sulfur-poor environment) and Venus' \ce{H2SO4-H2O} haze layer (a sulfur-rich environment) both give $w \approx 0.75$ \citep{Turco1979a,Russell1996,Seinfeld2012,Hansen1974,Ragent1985}. We thus set $w = 0.75$ for both limiting and best conditions.

Once nucleated, aerosols grow rapidly via condensation if either \ce{H2SO4} or \ce{H2O} gas is supersaturated 
\citep{Turco1979a,Seinfeld2012}. The aerosols additionally grow via coagulation:  diffusion and turbulence lead to sticking collisions between aerosols, which increase particle size and decrease particle number \citep[e.g.,][]{Seinfeld2012}. In both Earth's and Venus's atmospheres, \ce{H2SO4-H2O} aerosols tend to be relatively mono-dispersed in size \citep{Knollenberg1980,Seinfeld2012}, 
so describing their size distribution by a single average radius $r$ is a valid approximation.

We estimate a reasonable $r$ value from the \ce{H2SO4-H2O} aerosols that dominate Venus' haze layer's radiative properties as $r^\text{b}$ = 1 $\mu$m \citep{Hansen1974,Knollenberg1980}. From the inverse dependence of $\kappa_\text{e}$ on $r$ in Equation \eqref{eq:kappa}, smaller $r$ values are more favorable for haze formation for a given number of atmospheric sulfur atoms. However, particles too small will be observationally indistinguishable from Rayleigh scattering by gas molecules. We therefore set $r^\ell$ by considering where $Q_\text{e}$ transitions from the Rayleigh scattering limit to the Mie scattering regime. 

\begin{figure}
\centering
\plotone{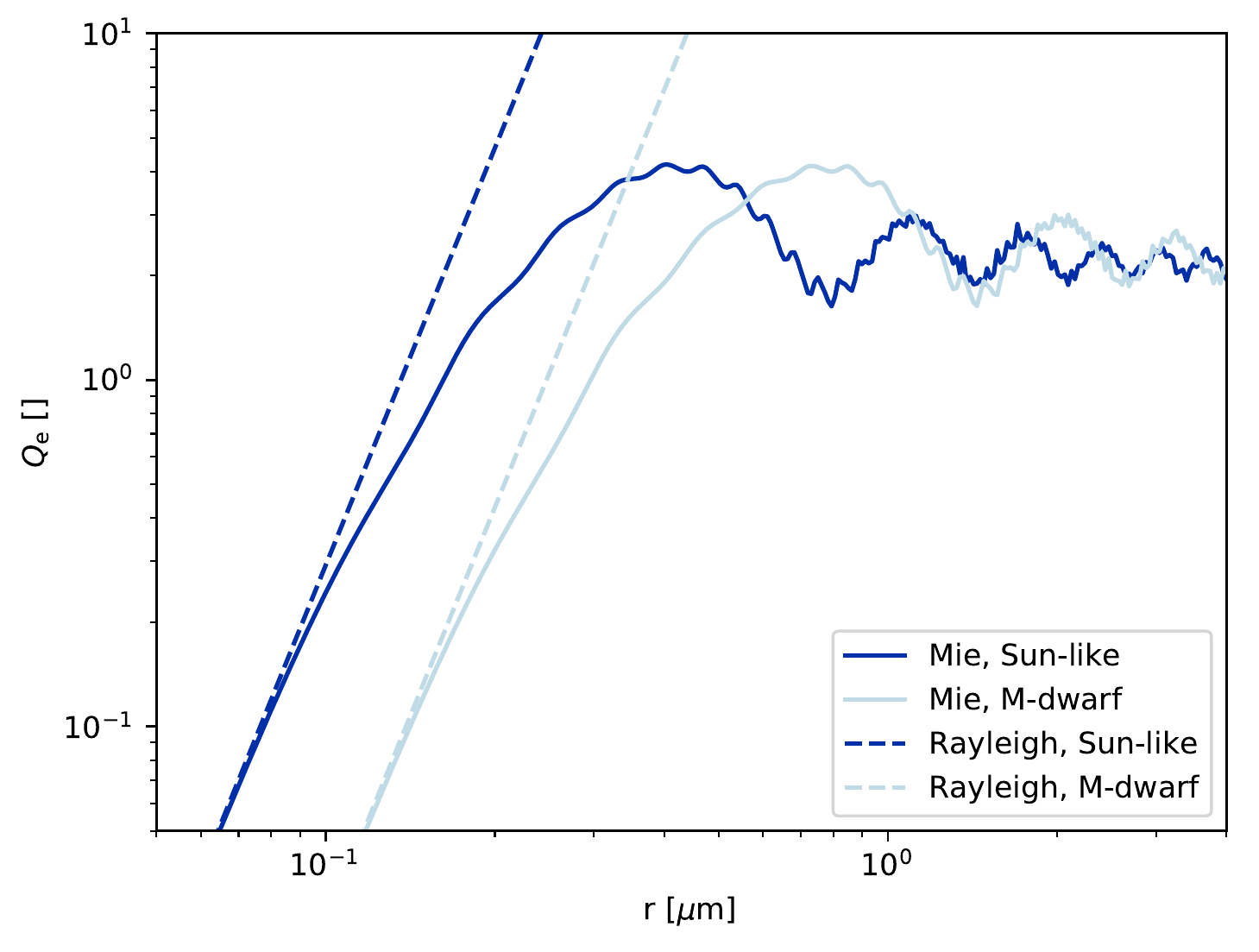}
\caption{Extinction efficiency ($Q_\text{e}$) of \ce{H2SO4-H2O} aerosols versus average particle radius ($r$). Solid lines are calculated from Mie theory and dashed lines show the Rayleigh limit (valid approximation for small particles). Values for a Sun-like star are in dark blue and those for an M dwarf in light blue.}
\label{fig:mie}
\end{figure}

$Q_\text{e}$ is calculated with Mie theory from $r$, the composition of the particle (within the \ce{H2SO4-H2O} system, specified by $w$), the index of refraction of this given \ce{H2SO4-H2O} mixture \citep{Palmer1975}, and the wavelength of incident light ($\lambda$) being attenuated \citep{Bohren2008}. We take $\lambda$ to be the peak of the planet's host star's spectrum, which is easily calculated from Wien's displacement law given the star's observed effective temperature. Figure \ref{fig:mie} shows $Q_\text{e}$ as a function of $r$ for both Sun-like ($\lambda = 0.556 \ \mu$m) and M-dwarf stars (for illustrative purposes, $\lambda = 1 \ \mu$m) with the Rayleigh limit superimposed. Based on this calculation, we set $r^\ell$ = 0.1 $\mu$m for a Sun-like star and $r^\ell$ = 0.2 $\mu$m for an M dwarf.  

We calculate the aerosol layer vertical path optical depth $\delta$ as 
\begin{equation} \label{eq:op_depth}
    \delta = u \kappa_\text{e}
\end{equation}
where $u$ is the mass column of aerosol particles. 
We estimate the critical minimum optical depth value for an observable haze layer $\delta^\ast$ by simulating transmission spectra. This procedure is described in Section \ref{subsec:obs_aero} after we finish outlining the structure of the rest of the atmosphere.

From the definition of mass column $u$, we can calculate the critical mass of \ce{H2SO4} $M^\ast_{\ce{H2SO4}}$ as
\begin{equation}\label{eq:crit_mass}
    M^\ast_{\ce{H2SO4}} = u^\ast w 4\pi R_\text{P}^2
\end{equation}
where $R_\text{P}$ is the radius of the planet. 
Putting $u^\ast$ in terms of $\delta^\ast$ and aerosol parameters with Equations \eqref{eq:kappa} and \eqref{eq:op_depth}, we then calculate the critical number of \ce{H2SO4} molecules in aerosols needed to create an observable haze layer as
\begin{equation}\label{eq:Nh2so4crit}
    N^\ast_{\ce{H2SO4}} = \frac{M^\ast_{\ce{H2SO4}}}{m_{\ce{H2SO4}}} = \frac{16 \pi}{3}\frac{\delta^\ast}{Q_\text{e}}r\rho_\text{aero} R_\text{P}^2\frac{w}{m_{\ce{H2SO4}}}
\end{equation}
where $m_{\ce{H2SO4}}$ is the mass of a molecule of \ce{H2SO4}.

\subsection{Aerosol Sedimentation and Mixing}
Once in the troposphere, \ce{H2SO4}-\ce{H2O} aerosols are ideal cloud condensation nuclei because of their high affinity for water, and they are rapidly removed by precipitation \citep{Seinfeld2012}.
For a mean residence time $\tau_\text{life}$, \ce{H2SO4} remains in the stratosphere as an aerosol contributing to the optical depth of the sulfur haze. We only consider \ce{H2SO4} as radiatively relevant until it is transported to the tropopause---the boundary between the isothermal stratosphere and convecting troposphere---because of \ce{H2SO4}'s short lifetime in the troposphere. 

The parameter $\tau_\text{life}$ depends on the size of the aerosol. If coagulation and condensation are effective and particles grow large enough, they will gravitationally settle out of the upper atmosphere. Otherwise, small particles will instead likely be removed by dynamic processes first due to their slow settling velocities. 
Thus, the average lifetime of an \ce{H2SO4-H2O} aerosol in the stratosphere is approximately
\begin{equation}
    \tau_\text{life} = \min \{\tau_\text{fall},\tau_\text{mix}\}
\end{equation}
where $\tau_\text{fall}$ is the gravitational settling timescale and $\tau_\text{mix}$ is the dynamic mixing timescale of exchange between the stratosphere and troposphere. 

We calculate $\tau_\text{fall}$ from the Stokes velocity of the falling particle as
\begin{equation}
    \tau_\text{fall} = \frac{z_\text{fall}}{v_\text{Stokes}} = \frac{9}{2}\frac{z_\text{fall}\eta}{r^2 g \rho_\text{aero} C_\text{C}}.
\end{equation}
Here, $g$ is planet surface gravity, $\eta$ is the dynamic viscosity of air, and $C_\text{C}$ is the Cunningham-Stokes correction factor for drag on small particles \citep{Seinfeld2012}. $z_\text{fall}$ is the average distance a sulfur aerosol must fall from its formation location to the tropopause. We conservatively estimate this parameter as the scale height of the stratosphere (i.e., $z_\text{fall} = RT_\text{strat}(\mu_\text{air}g)^{-1}$).

Calculating $\tau_\text{mix}$ from first principles is not possible because a general theory of stratosphere-troposphere mixing timescales as a function of external parameters does not yet exist. However, insights can be gained from studying Earth. Today, Earth's strong stratospheric temperature inversion inhibits this stratosphere-troposphere exchange. Since radiatively generating an upper atmospheric temperature inversion stronger than Earth's is difficult,
Earth's $\tau_\text{mix}$ is likely near an upper bound on stratosphere-troposphere mixing timescales. On the most observable M-dwarf planets, permanent day-night sides of planets due to tidal effects are predicated to generate strong winds \citep[e.g.,][]{Showman2010,Showman2011} that will almost certainly reduce $\tau_\text{mix}$ from Earth values. Therefore, from Earth-based timescales, we estimate $\tau_\text{mix}$ = 1 year for all cases \citep{Warneck1999}.

\subsection{Aerosol Formation} \label{subsec:photochemistry}
The saturation vapor pressure of \ce{H2SO4} at expected terrestrial planet stratosphere temperatures is extremely low ($\ll 10^{-6}$ Pa) \citep{Kulmala1990}, so all stratospheric \ce{H2SO4} is effectively found condensed in aerosols rather than as a gas. The sulfur source for \ce{H2SO4-H2O} aerosols in oxidized atmospheres is \ce{SO2} gas. Here, we define $\tau_{\ce{SO2}\rightarrow \ce{H2SO4}}$ as the timescale required to convert stratospheric \ce{SO2} to \ce{H2SO4}.  $\tau_{\ce{SO2}\rightarrow \ce{H2SO4}}$ is challenging to calculate precisely because \ce{SO2} is oxidized to \ce{H2SO4} through a series of photochemical reactions that are poorly constrained both in terms of reaction pathways and rates \citep[][etc.]{Turco1979a,Stockwell1983,Yung1998,Burkholder2015}. The two main stages of the conversion of \ce{SO2} to \ce{H2SO4} are (1) \ce{SO2} is converted to \ce{SO3} via a photochemical product and (2) \ce{SO3} and \ce{H2O} react to form \ce{H2SO4}
\citep{Burkholder2015}.
For part 1 of the conversion, commonly proposed reactions are 
\begin{equation}
    \ce{SO2 + O + M -> SO3 + M}
\end{equation}
\citep{Yung1998,Burkholder2015} or 
\begin{equation}
    \ce{SO2 + OH + M -> HSO3 + M}
 \end{equation}
 \citep{Turco1979a,Stockwell1983,Burkholder2015} followed by either 
 \begin{align}
     \ce{HSO3 + O2 &-> SO3 + HO2}\\
     \ce{HSO3 + OH &-> SO3 + H2O}
 \end{align}
 \citep{Turco1979a,Stockwell1983,Burkholder2015}. For part 2, the proposed reaction is
 \begin{equation}
     \ce{SO3  + H2O -> H2SO4},
 \end{equation}
but it is unclear whether this reaction proceeds as a two-body or three-body reaction kinetically \citep{Seinfeld2012,Burkholder2015}.  Large uncertainties in reaction rates for these reactions are compounded by the reaction kinetics' sensitivity to photochemical product (\ce{OH}, \ce{O}) concentrations, which require detailed knowledge of atmospheric composition for accurate results \citep{Jacob1999}. 

Despite these uncertainties, we can make simplifications because \ce{SO2}, \ce{H2O}, and stellar UV photons are essential for the conversion, regardless of the pathway. In particular, because photochemistry is the rate limiting step except potentially in extremely dry atmospheres, we can estimate a lower limit of $\tau_{\ce{SO2}\rightarrow \ce{H2SO4}}$ by considering how long it takes for the number of UV photons necessary for the reaction to proceed to strike the planet's atmosphere. 

For photolysis-driven reactions, the lifetime $\tau$ of the photolyzed species can be described by
\begin{equation}\label{eq:lifetime}
    \tau^{-1} = \frac{1}{4}\int\limits_{\lambda_\text{UV}} q(\lambda) \sigma (\lambda) \phi_\gamma(\lambda) \mathrm{d} \lambda
\end{equation}
where $\lambda$ is wavelength of light, $q$ is the quantum yield or the probability that a photon will dissociate a molecule, $\sigma$ is the absorption cross section of the molecule, $\phi_\gamma$ is the flux of photons to a planet's atmosphere per unit wavelength, and the factor of $\sfrac{1}{4}$ accounts for day–night and stellar zenith angle averaging \citep{Jacob1999,Cronin2014}. 
To represent the most rapid conversion rate of \ce{SO2} to \ce{H2SO4}, we set $q = 1$ for all wavelengths (i.e., 100\% probability that a photon will dissociate a molecule it strikes). 
At minimum, the \ce{SO2} to \ce{H2SO4} reaction requires one \ce{O} or \ce{OH} molecule. For weakly to strongly oxidized terrestrial planet atmospheres, this photochemical product will most likely be the result of the dissociation of \ce{H2O}, \ce{O2}, or \ce{CO2}. (Ostensibly, photodissociated \ce{SO2} could also provide \ce{O}, but at the surface number densities of \ce{SO2} we consider here, \ce{SO2} is so optically thin that it cannot effectively absorb photons \citep{Manatt1993,Hamdy1991}.)  At each $\lambda$, we set $\sigma (\lambda)$ as the maximum absorption cross section from these three molecules, using data given in \cite{Chan1993}, \cite{Mota2005}, \cite{Lu2010}, \cite{Yoshino1988}, \cite{Kockarts1976}, \cite{Brion1979}, and \cite{Huestis2010}. 

To set the wavelength for the upper limit of integration, we take the smallest minimum bond dissociation energy of \ce{H2O}, \ce{O2}, and \ce{CO2} and convert it to a wavelength (via $E_\text{dis} = E_\gamma = h\nu = h c / \lambda$). A minimum energy of $8.2 \times 10^{-19}$ J \citep{deB1970} yields a maximum wavelength of 240 nm. We begin the integration from $\lambda = 0$ nm for a maximum number of photons. 
In general, $\phi_\gamma$ will be determined from the host star's spectrum and the distance of the planet from its host star. 
Here, we use the present-day flux of the Sun to Earth as measured by \cite{Thuillier2004} as a baseline $\phi_\gamma$ for a Sun-like star. The ratio of UV flux to stellar bolometric flux declines as stellar mass declines (due to lower effective temperatures and thus higher wavelength Planck peaks)
while the ratio of extreme UV (EUV) flux to bolometric flux tends to increase for M dwarfs relative to Sun-like stars due to higher magnetic activity \citep{Scalo2007,Shkolnik2014,Schaefer2016,Wordsworth2018}.
The precise change in UV and EUV flux relative to bolometric flux for an M dwarf relative to the Sun will depend on the specific star and its age, but for illustrative purposes here, we assume a $10 \times$ decrease in UV flux and a $10 \times$ increase in EUV flux relative to the solar value. (Here, we define the EUV wavelength cutoff as $\lambda = 91$ nm, following \cite{Wordsworth2018}.) Plugging in values to Equation \eqref{eq:lifetime}
then leads to $\tau_{\ce{SO2}\rightarrow \ce{H2SO4}}^\ell \approx$ 3.4 days for a Sun-like star and $\tau_{\ce{SO2}\rightarrow \ce{H2SO4}}^\ell \approx$ 2.5 days for an M dwarf. On modern Earth, $\tau_{\ce{SO2}\rightarrow \ce{H2SO4}} \approx 30$ days \citep{Turco1979b,Macdonald2017}, which we take as $\tau_{\ce{SO2}\rightarrow \ce{H2SO4}}^\text{b}$.

Once we have estimated the lifetimes of \ce{H2SO4} and \ce{SO2} in the stratosphere, we can directly relate the  steady-state $N_{\ce{H2SO4}}$ in aerosols to $N_{\ce{SO2}}$ in gas by assuming that the production rate of \ce{H2SO4} is equal to its removal rate:
\begin{equation}\label{eq:NH2SO4}
    \frac{N_{\ce{H2SO4}}}{\tau_\text{life}} = \frac{N_{\ce{SO2}}}{\tau_{\ce{SO2}\rightarrow \ce{H2SO4}}}.
\end{equation}
From Equation \eqref{eq:NH2SO4} and the definition of pressure as force over area, $N_{\ce{SO2}}^\ast$ converted to a critical partial pressure of \ce{SO2} ($p_{\ce{SO2}}^\ast$) at the tropopause is 
\begin{equation}
    p_{\ce{SO2},\text{tropopause}}^\ast = N^\ast_{\ce{H2SO4}} \frac{g m_{\ce{SO2}}}{4\pi R_\text{P}^2} \frac{
    \tau_{\ce{SO2}\rightarrow \ce{H2SO4}}}{\tau_\text{life}}  
\end{equation}
where $m_{\ce{SO2}}$ is the mass of a molecule of \ce{SO2}.

\subsection{Lower Atmosphere Transport}\label{subsec:loweratm}
On a wet, temperate planet, \ce{SO2} faces no significant sinks below the photochemically active upper atmosphere until it encounters the water cloud layer in the troposphere. Once \ce{SO2} encounters condensed water (as either rain or cloud droplets), the \ce{SO2} dissolves and is subsequently rained out of the atmosphere \citep{Giorgi1985,Seinfeld2012,Hu2013}. The high effective solubility of \ce{SO2} means that effectively all \ce{SO2} at and below the cloud deck is removed every rainfall event \citep{Giorgi1985}. This process of \emph{wet deposition} greatly limits the lifetime of \ce{SO2} in the lower atmosphere and decreases the mixing ratio of \ce{SO2} ($f_{\ce{SO2}} \equiv p_{\ce{SO2}} / p$) beyond the water cloud layer in the upper atmosphere relative to the surface. 

In our model, we take the mixing ratio of \ce{SO2} at the tropopause to be directly proportional to the mixing ratio of \ce{SO2} at the surface, such that
\begin{equation} \label{eq:pso2surf_aero}
    \frac{p_{\ce{SO2},\text{tropopause}}}{p_\text{tropopause}} = \alpha \frac{p_{\ce{SO2},\text{surf}}}{p_\text{surf}}.
\end{equation}
Here, $p$ is the total atmospheric pressure, and $\alpha$ is the change of mixing ratio between the surface and the upper atmosphere due to the effects of wet deposition. We do not yet know much about hydrological cycles on planets with less surface liquid water than Earth, so in the limiting case we completely ignore wet deposition and set $\alpha^\ell$ = 1.  For the reasonable case, we assume $\alpha^\text{b}$ = 0.1, based on comparison with \ce{SO2} vertical profiles measured on Earth \citep{Georgii1978,Meixner1984}. 

We calculate a pressure-temperature profile from surface temperature $T_\text{surf}$ and pressure $p_\text{surf}$ assuming a dry adiabat until water vapor becomes saturated and then a moist adiabat until a specified (isothermal) stratospheric temperature $T_\text{strat}$ is reached, following the derivation of \cite{Wordsworth2013}. This calculation requires an assumed mixing ratio of water $f_{\ce{H2O}} \equiv p_{\ce{H2O}} / p$ at the surface. This mixing ratio is calculated from relative humidity at the surface $\text{RH}_\text{surf}$ and $T_\text{surf}$ via $f_{\ce{H2O}\text{,surf}} = \text{RH}_\text{surf}\times p_\text{sat,\ce{H2O}}(T_\text{surf})$ where $p_\text{sat,\ce{H2O}}(T_\text{surf})$ is the saturation pressure of water at temperature $T_\text{surf}$. We set $\text{RH}_{\text{surf}}^\text{b}$ = 0.77 like the Earth. Increasing $f_{\ce{H2O}}$ yields a higher tropopause, which results in a higher $p_{\ce{SO2},\text{surf}}$ for a given $p_{\ce{SO2},\text{tropopause}}$. Therefore, we set $\text{RH}_{\text{surf}}^\ell$ = 0 (yielding a dry adiabat atmospheric structure).

\subsection{Aerosol Observability}\label{subsec:obs_aero}
Having reviewed the atmospheric conditions dictated by the presence of an observable \ce{H2SO4-H2O} haze layer, we can now return to the problem of how to calculate the critical optical depth $\delta^\ast$ for observation. We simulate the transmission spectra expected of a planet with a sulfur haze for a range of optical depths to determine at which $\delta$-value the hazy spectrum becomes distinct from the clear spectrum.
The appendices of \cite{Morley2015,Morley2017} provide the details of our model for simulating transmission spectra, which uses the matrix prescription presented in \cite{Robinson2017}. We calculate molecular cross sections as described in \cite{Freedman2008,Freedman2014}, including an updated water line list from \cite{Polyansky2018}.

As inputs to our model, we require aerosol size and number densities, atmospheric gases number densities, and temperature as functions of pressure. From Equation \eqref{eq:NH2SO4}, we calculate an \ce{H2SO4-H2O} aerosol number density profile, assuming exponential decay in aerosols from the tropopause until a parameterized cutoff height (due to lack of photochemical \ce{H2SO4} production).
We input atmospheric number densities assuming constant mixing ratios (except for \ce{H2O}) for a given atmospheric composition. The temperature-pressure profile follows a moist adiabat until it reaches an isothermal stratosphere. The mixing ratio of \ce{H2O} $f_{\ce{H2O}}$ is determined by the moist adiabat. $f_{\ce{H2O}}$ is set at the surface and is held constant until water vapor becomes saturated. Then, $f_{\ce{H2O}}$ evolves according to water's saturation pressure until the tropopause, above which $f_{\ce{H2O}}$ remains constant. 

Varying $\delta$ in these simulated transit spectra for an Earth-like atmosphere (see Section \ref{subsec:res_obs_haze} and Figure \ref{fig:spectra})
suggested that the critical minimum value for an observable haze layer is $\delta^\ast \approx 0.1$. Different atmospheres will yield different $\delta^\ast$ values, but likely not by orders of magnitude. The precise shape of a hazy planet's transit spectrum is also somewhat sensitive to the photochemical haze cutoff height. Lower cutoff heights yield more traditionally haze-characteristic flat spectra \citep[e.g.,][]{Kreidberg2014}, and higher, less physical cutoff heights yield more structured---though still smoothed relative to a clear atmosphere---spectra; but the full range of possible haze cutoff values still yields spectra identifiable as hazy. In the absence of a detailed photochemical model, we choose to display results in Figure \ref{fig:spectra} with a cutoff height of two stratospheric scale heights (11.7 km), following Earth and Venus \ce{H2SO4-H2O} aerosol profile observations \citep[e.g.,][]{Sekiya2016,Knollenberg1980}.

\subsection{Ocean Sulfur Storage} \label{subsec:atmoc}
When surface liquid water is present, convection and the hydrological cycle act to bring the lower atmospheric and dissolved sulfur in the ocean into equilibrium rapidly. The partial pressure of \ce{SO2} at the surface ($p_{\ce{SO2},\text{surf}}$) 
is held in equilibrium with the concentration of dissolved, aqueous \ce{SO2} in the ocean [\ce{\ce{SO2} (aq)}] via Henry's law: 
\begin{equation}\label{eq:Henry}
    p_{\ce{SO2},\text{surf}} = K_\text{H}(T) [\ce{\ce{SO2} (aq)}],
\end{equation}
where $K_H = 6.96 \times 10^4$ Pa L mol$^{-1}$ is the Henry's law constant for \ce{SO2} \citep{Pierrehumbert2010}.
Once dissolved, \ce{SO2} reacts with the ambient water to form sulfurous acid, which dissociates to \ce{H+}, \ce{HSO3-}, and \ce{SO3^2-} ions:
\begin{align}
    \ce{SO2 (aq) + H2O &<=> HSO3- + H+}\label{eq:aq_react1}\\ 
    \ce{HSO3- &<=> SO_3^2- + H+} \label{eq:aq_react2}
\end{align}
\citep{Neta1985,Halevy2007}. The ocean's pH (log of the concentration of \ce{H+} ions) controls the partitioning of sulfur between \ce{SO2(aq)}, \ce{HSO3-}, and \ce{SO3^2-}---species which are collectively known as S(IV) or sulfur in a +4 redox state. As we discuss in the next section, however, the picture of aqueous sulfur chemistry given by Equations \eqref{eq:Henry}-\eqref{eq:aq_react2} is incomplete because S(IV) disproportionation is effective under a wide range of conditions. We return to this point shortly.

The atmosphere is only in equilibrium with \ce{SO2(aq)}, so when Equations \eqref{eq:Henry}-\eqref{eq:aq_react2} hold,  the ocean's pH regulates its ability to store sulfur versus the atmosphere. Ocean pH is treated as an independent variable in our model because it has a strong influence on the sulfur distribution and is difficult to estimate from first principles \citep[e.g.,][]{Kempe1985,Macleod1994,Sleep2001,Halevy2017}. Our model's pH dependence is discussed further in Section \ref{subsec:obs_sulfur_oc}.

With the partial pressure of \ce{SO2} at the surface and an assumed pH, Equations \eqref{eq:Henry}-\eqref{eq:aq_react2} 
give the concentration of S(IV) species [S(IV)(aq)] in the ocean as
\begin{align}\label{eq:siv_oc}
    [\ce{S(IV)(aq)}] &= [\ce{SO2(aq)}] + [\ce{HSO3-}] + [\ce{SO3^2-}] \nonumber \\ &=  \frac{p_{\ce{SO2},\text{surf}}}{K_H} \left ( 1 + \frac{K_1}{[\ce{H+}]} + \frac{K_1 K_2}{[\ce{H+}]^2} \right ).
\end{align}
Here [\ce{H+}] = $10^{-\text{pH}}$, and $K_1 = 10^{-1.86}$ and $K_2 = 10^{-7.2}$ are the first and second, respectively, acid dissociation constants of sulfurous acid \citep{Neta1985}. These constants depend on salinity, pressure, and temperature in general, although we neglect this dependence here.

At this point, we incorporate the most significant variable in our model:
the planet's total mass of surface liquid water $M_\text{oc}$. Like pH, $M_\text{oc}$ is taken to be an independent variable because it is the unknown we are interested in investigating. The total number of sulfur atoms needed in the atmosphere-ocean system to achieve observable atmospheric sulfur $N_{\ce{S}}^\ast$ is a function of [S(IV)(aq)], $p_{\ce{SO2},\text{surf}}$, and $M_\text{oc}$. It can be written as the sum of sulfur atoms required in the atmosphere and the ocean:
\begin{equation} \label{eq:Scrit}
    N_{\ce{S}}^\ast = N_{\ce{S},\text{atm}}^\ast + N_{\ce{S(IV)},\text{oc}}^\ast.
\end{equation}
To avoid having to prescribe an $f_{\ce{SO2}}$ profile, we make the simplifying assumption that \ce{SO2} is always well mixed in calculating $N_{\ce{S},\text{atm}}^\ast$. For an observable level of \ce{SO2}, the critical number of sulfur atoms in the atmosphere $N_{\ce{S},\text{atm}}^\ast$ is
\begin{equation}\label{eq:S_atm_crit_so2}
    N_{\ce{S},\text{atm}}^\ast = \frac{4\pi R_\text{P}^2}{g} p_{\ce{SO2},\text{surf}}^\ast \frac{N_A}{\mu_{\ce{SO2}}}
\end{equation}
where $\mu_{\ce{SO2}}$ is the molar mass of \ce{SO2} and $p_{\ce{SO2},\text{surf}}^\ast$ is given by Equation \eqref{eq:pso2surf_so2};
whereas for an observable haze layer
\begin{equation}\label{eq:S_atm_crit_haze}
    N_{\ce{S},\text{atm}}^\ast = \frac{4\pi R_\text{P}^2}{g} p_{\ce{SO2},\text{surf}}^\ast \frac{N_A}{\mu_{\ce{SO2}}} + N_{\ce{H2SO4}}^\ast
\end{equation}
where $p_{\ce{SO2},\text{surf}}^\ast$ is given by Equation \eqref{eq:pso2surf_aero} and $N_{\ce{H2SO4}}^\ast$ by Equation \eqref{eq:Nh2so4crit}. For both observable atmospheric sulfur products, the critical number of S(IV) sulfur atoms in the ocean $N_{\ce{S(IV)},\text{oc}}^\ast$ is 
\begin{equation}\label{eq:S_crit_ocean}
    N_{\ce{S(IV)},\text{oc}}^\ast = [\ce{S(IV)(aq)}]^\ast \frac{M_\text{oc}}{\rho_{\ce{H2O}}} \left ( 1000 \ \text{L}\ \text{m}^{-3}\right )N_A
\end{equation}
where $N_A$ is Avogadro's number, $[\ce{S(IV)(aq)}]^\ast$ is given by Equation \eqref{eq:siv_oc}, and the factor of 1000 L m$^{-3}$ converts the moles per L of $[\ce{S(IV)(aq)}]^\ast$ to SI units. 

\subsection{Expected Sulfur in Surface Reservoir}
We now need to calculate the number of sulfur atoms expected in the atmosphere and ocean to compare to the critical value given by Equation \eqref{eq:Scrit} to determine whether observable \ce{SO2} buildup or haze formation is reasonable. This calculation is simplified because aqueous \ce{HSO3-} and/or \ce{SO3^2-} are not thermodynamically stable  \citep{Karchmer1970,Hayon1972,Brimblecombe1989,Guekezian1997,Jacobson2000,Ermakov2001,Halevy2013}.
Spontaneous or easily catalyzed disproportionation and oxidation reactions convert sulfur from \ce{S(IV)} to \ce{S(VI)} and/or \ce{S(0)} \citep{Karchmer1970,Guekezian1997,Johnston2011,Halevy2013}, the former even in the absence of dissolved oxygen \citep{Guekezian1997,Zopfi2004}. The proposed stoichiometric oxidation reaction is
\begin{align}
    \ce{2SO3^2- + O2 &-> 2SO4^2-}
\end{align}
though the chain of reactions from which this reaction proceeds is poorly understood \citep{Ermakov2001,Halevy2013}.
Proposed stoichiometric disproportionation reactions in anoxic waters include:
\begin{align}
    \ce{3H2SO3 &-> 2H2SO4 + S + H2O} \\
    \ce{7SO3^2- + 3H2O & -> S4O6^2- + 3SO4^2- + 6OH^-}\\
    \ce{4SO3^2- + H2O &->  2SO4^2- + S2O3^2- + 2OH^-}\\
    \ce{3SO2 + 2H2O &-> 2 SO4^2- + S + 4 H+}
\end{align}
\citep{Karchmer1970,Guekezian1997,Ryabinina1972}.

 These disproportionation reactions are thermodynamically favored based on their calculated Gibbs free energy changes \citep{Guekezian1997}.  However, the precise pathways by which S(IV) decay proceeds and their respective reaction rates are unclear, even in environments of well-known pH, [\ce{O2(aq)}], and $T$ \citep{Ermakov2001}. Nonetheless, experiments suggest across wide pH, [\ce{O2(aq)}], and $T$ ranges that \ce{HSO3-} and \ce{SO3^2-} are unstable on rapid timescales of seconds to weeks \citep{Ermakov2001,Avrahami1968,Guekezian1997,Suzuki1999,Zopfi2004}. 
Indeed, in Earth's modern oceans, [\ce{HSO3-}] and [\ce{SO3^2-}] concentrations are close to zero, with \ce{SO4^2-} ions (S(VI)) being the only major dissolved sulfur constituent, despite the fact that \ce{SO2} is the dominant sulfur outgassing product \citep{Schlesinger2013}. For Earth's pH and average surface $p_{\ce{SO2}}$, the expected [\ce{SO3^2-}] concentration levels in the ocean  in the absence of disproportionation or oxidation are of order 0.01-0.001 mol L$^{-1}$. Measurements of [\ce{SO3^2-}] are of order $1 \times 10^{-6}$~mol L$^{-1}$ or less \citep{Goldhaber2003}. 

After the oxidation or disproportionation of \ce{S(IV)}, both possible sulfur products \ce{S(0)} and \ce{S(VI)} are lost to the reservoir of sulfur in equilibrium with the atmosphere until they have been reprocessed by the interior. \ce{S(0)} is insoluble \citep{Boulegue1978}, so its formation removes that sulfur from the aqueous reservoir. 
\ce{S(IV)} is soluble, but there are no known abiotic pathways to reduce S(VI) back to S(IV) \citep{Brimblecombe1989}. We discuss the implications of the presence of sulfur-consuming life on the sulfur cycle in Section \ref{subsec:life}, but it is not expected to substantially alter this picture. 

Theoretically, aqueous S(VI) could be a source of \ce{H2SO4} gas from Henry's law. However, because \ce{H2SO4} is an extremely strong acid, dissolved \ce{H2SO4} will not be present as \ce{H2SO4(aq)}. Significant \ce{H2SO4(aq)} requires physically impossible water pHs $\lesssim$ -3 \citep{Jacobson2000}. Thus, dissolved S(VI) species cannot contribute to atmospheric \ce{H2SO4} either. 

As neither S(0) nor S(VI) is in equilibrium with atmospheric \ce{SO2}, aqueous chemistry can act as a pump for rapidly removing \ce{SO2} from the atmosphere, and there is essentially no potential for significant atmospheric sulfur gas buildup over geologic time on planets with surface liquid water \citep[e.g.,][]{Kasting1989}. A central outcome of our analysis is, therefore, that the sulfur outgassing flux must continuously counterbalance aqueous S(IV) decay if significant \ce{SO2} is to be present in an atmosphere. 

To see whether the buildup of observable \ce{SO2} or the formation of an observable haze layer is reasonable, we can compare the critical number of sulfur atoms in the atmosphere and ocean required for observation ($N_{\ce{S}}^\ast$) with the expected total number of sulfur atoms in the planet's ocean and atmosphere ($N_{\ce{S}}$). As the sulfur must be supplied by recent outgassing, we estimate $N_{\ce{S}}$ as 
\begin{equation}
    N_{\ce{S}} = \dot{N}_{\ce{S}} \tau_{\ce{S(IV)}}
\end{equation}
where $\dot{N}_{\ce{S}}$ is the global outgassing rate of sulfur (number of S atoms outgassed per unit time) and $\tau_{\ce{S(IV)}}$ is the timescale for \ce{S(IV)} to decay in the ocean. In order for a sulfur haze layer to be observed, we must have 
\begin{equation} \label{eq:hazeform}
    N_{\ce{S}} / N_{\ce{S}}^\ast \geq 1.  
\end{equation}

To determine the most favorable conditions for atmospheric sulfur buildup, we must evaluate a maximum sulfur outgassing rate $\dot{N}_{\ce{S}}$. 
This estimation is hampered by the lack of a clear path to placing a reasonable upper bound given large uncertainties in the quantitative chemistry that governs the transport of sulfur from the planetary interior to the surface as well as a large planetary parameter space; we make a good faith effort to estimate this maximum value but highlight this calculation as a clear region for future study. We do note that, given our interest in an equilibrium state, we are considering here maximum time-averaged outgassing rates rather than an instantaneous values from transient extreme events.

The precise frequency and magnitude of the outgassing events that determine $\dot{N}_{\ce{S}}$ will be a function of a planet's tectonic regime, which at present is extremely poorly constrained for exoplanets, given conflicting results from interior dynamics models \citep[e.g.,][]{oneill2007,Valencia2007,Stamenkovic2012}. 
Nonetheless, \cite{Kite2009} estimate theoretical upper bounds on terrestrial planet outgassing for both plate tectonics and stagnant lid regimes. They find outgassing rate is a strong decreasing function of planet age after $\sim$1 Gyr and roughly constant across bodies of different sizes when normalized per unit planet mass. In their model, outgassing peaks at about $20 \times$ modern-Earth outgassing levels. 

To translate an upper bound on total outgassing rate into an upper bound on sulfur outgassing rate, we assume that outgassing rate is proportional to melt production rate, following \cite{Kite2009}. 
For near (or above) Earth-like bulk sulfur contents, sulfur directly dissolved in melt is largely limited by the saturation limit of the sulfur-bearing species in the melt \citep[e.g.,][]{Anderson1974,Mathez1976,Edmonds2017} rather than availability of ambient interior sulfur. Experimentally, the sulfur melt saturation limit is most strongly controlled by melt oxidation state and then temperature with smaller effects from melt composition and pressure \citep{Jugo2005a,Jugo2009,Righter2009}.

Sulfur directly dissolved in melt is further in equilibrium with a volatile vapor phase also held within the melt \citep[e.g.,][]{Zajacz2012}. This vapor phase is what outgasses. The partitioning of sulfur between melt and vapor is a function of temperature, pressure, oxidation state, and melt composition \citep{Shinohara2008,Zajacz2012}. Recent experimental studies suggest that the dependence on melt oxidation state of this vapor-melt partitioning and of dissolved melt sulfur saturation counteract each other such that the total gaseous sulfur available for outgassing is essentially independent of interior oxidation state \citep{Jugo2005a,Zajacz2012}. The coupling of other dependencies between vapor-melt partitioning and melt saturation is less clear, but qualitatively these dependencies tend to (like oxidation state) act in opposition---muting, rather than amplifying, the effect of a given state variable on total sulfur outgassing potential. 

Melts tend to outgas at the surface because decreasing pressure increases the partitioning of sulfur into vapor relative to the melt.
Traditionally it has been assumed that initial sulfur vapor content at melt formation (high pressure) is zero, and all sulfur ultimately outgassed originates from direct dissolution into melt at formation; however, more recent observations call into question this latter assumption \citep[see][and references therein]{Oppenheimer2011}. 
Sulfur budget allocating estimates from larger explosive eruptions with satellite-quantifiable gas plumes suggest that an ambient vapor reservoir can interact with the melt during its ascent and contribute more sulfur   \citep[e.g.,][]{Wallace1994,Gerlach1996,Shinohara2008,Oppenheimer2011}. The origins of this ambient vapor reservoir are not well understood \citep{Oppenheimer2011}. However, proposed hypotheses seem to require either (1) an older planet with a buildup of sulfur-rich sediments from the surface having been transported deeper into the interior, suggesting that the contributions of this ambient vapor reservoir should be muted for young planets with the highest outgassing fluxes or (2) a long-suppressed regional eruption, suggesting limited influence over a time-averaged outgassing rate \citep{Oppenheimer2011}.

From these considerations, we believe assuming that modern Earth's average sulfur outgassing per unit melt is within an order of magnitude of an upper bound is a reasonable assumption. The conclusion seems consistent with Earth history:  models of the Archean sulfur cycle suggest that, in order to reproduce observed sulfur isotope fractionations, the Archean sulfur outgassing rate was a fraction of the modern rate despite elevated total volcanism relative to present day \citep[e.g.,][]{Harman2018a}.
From the average current Earth sulfur outgassing rate ($\dot{N}_{\ce{S},\oplus}$), we can then estimate our limiting value for this sulfur outgassing rate parameter as 
\begin{equation}
    \dot{N}_{\ce{S}}^\ell = 10 \times 20 \dot{N}_{\ce{S},\oplus}\frac{M_p}{M_\oplus}
\end{equation}
where $\dot{N}_{\ce{S},\oplus}$ = $1.89 \times 10^{35}$ atoms S yr$^{-1}$  \citep{Schlesinger2013,Halmer2002}, $M_p$ is planet mass, $M_\oplus$ is Earth's mass, the factor of 20 is from \cite{Kite2009}'s calculations to account for the potential for increased total outgassing, and the additional factor of 10 is to account for the potential for higher sulfur content in melts than modern mass-averaged Earth values (e.g., from higher melt formation temperatures, greater influence of an at-depth vapor reservoir). 
We set the reasonable scenario from the modern-Earth outgassing value with 
\begin{equation}
    \dot{N}_{\ce{S}}^\text{b} =  \dot{N}_{\ce{S},\oplus}\frac{M_p}{M_\oplus}.
\end{equation}

The remaining key parameter in our model is the S(IV) aqueous decay timescale $\tau_{\ce{S(IV)}}$. As we have discussed, current understanding of the kinetics of aqueous S(IV) instability is still quite limited. Therefore, we solve directly for the critical $\tau_{\ce{S(IV)}}^\ast$ for observable sulfur buildup from a simplification of the condition for observation given by Equation \eqref{eq:hazeform}: 
\begin{equation}
    \tau^\ast_{\ce{S(IV)}} = N_{\ce{S(IV)},\text{oc}}^\ast/\dot{N}_{\ce{S}}.
\end{equation}
As $\tau_{\ce{S(IV)}}$ relates only to the destruction rate of aqueous \ce{S(IV)}, we make this conservative assumption---to neglect the direct contribution of atmospheric sulfur to the critical amount of sulfur necessary for atmospheric observation---in lieu of adding the exchange rate of \ce{S(IV)} between the ocean and atmosphere to our equilibrium model.  
This simplification is favorable to atmospheric sulfur buildup, underestimating critical sulfur required for observation in water-poor, low-pH regimes where $N_{\ce{S(IV)},\text{oc}} \ll N_{\ce{S},\text{atm}}$. 
As we will show, the variation in $\tau_{\ce{S(IV)}}^\ast$ value as a function of pH and ocean volume is so large that we are able to reach strong conclusions, even given the uncertainties.

\begin{deluxetable*}{cccc}
\tablecaption{Model inputs \label{tab:parameters}} 
\tablehead{\colhead{parameter} & \colhead{limiting value - Sun-like (M dwarf)} & \colhead{best estimate} & \colhead{detection method}}
\startdata
$u_{\ce{SO2},\text{surf}}^\ast$  [kg m$^{-2}$] & $2.3 \times 10^{-4}$ & $2.3 \times 10^{-2}$ & gas \\
$r_\text{avg}$ [$\mu$m]  & 0.1  (0.2) & 1 & aerosol \\ 
$w_{\ce{H2SO4}}$ [kg kg$^{-1}$]& 0.75  & 0.75 & aerosol\\
$\tau_{\ce{SO2}\rightarrow \ce{H2SO4}}$ [day] & 3.4 (2.5) & 30 & aerosol\\
$\tau_\text{mix}$ [yr] & 1 & 1 & aerosol\\
$f_{\ce{SO2},\text{tropopause}}/f_{\ce{SO2}\text{,surf}}$ [ ] & 1 & 0.1 & aerosol\\
$\dot{N}_{\ce{S}}$ [kg S yr$^{-1}$] & 200$\dot{N}_{\ce{S},\oplus}\frac{M_\text{P}}{M_\oplus}$ & 1$\dot{N}_{\ce{S},\oplus}\frac{M_\text{P}}{M_\oplus}$ & aerosol \& gas
\enddata
\tablenotetext{}{Parameters for which we plot results in Section \ref{sec:results}.}
\end{deluxetable*}

\section{Results}\label{sec:results}
\subsection{Detecting a Haze Layer}\label{subsec:res_obs_haze}
\begin{figure*}
\centering
\includegraphics[width=0.9\textwidth]{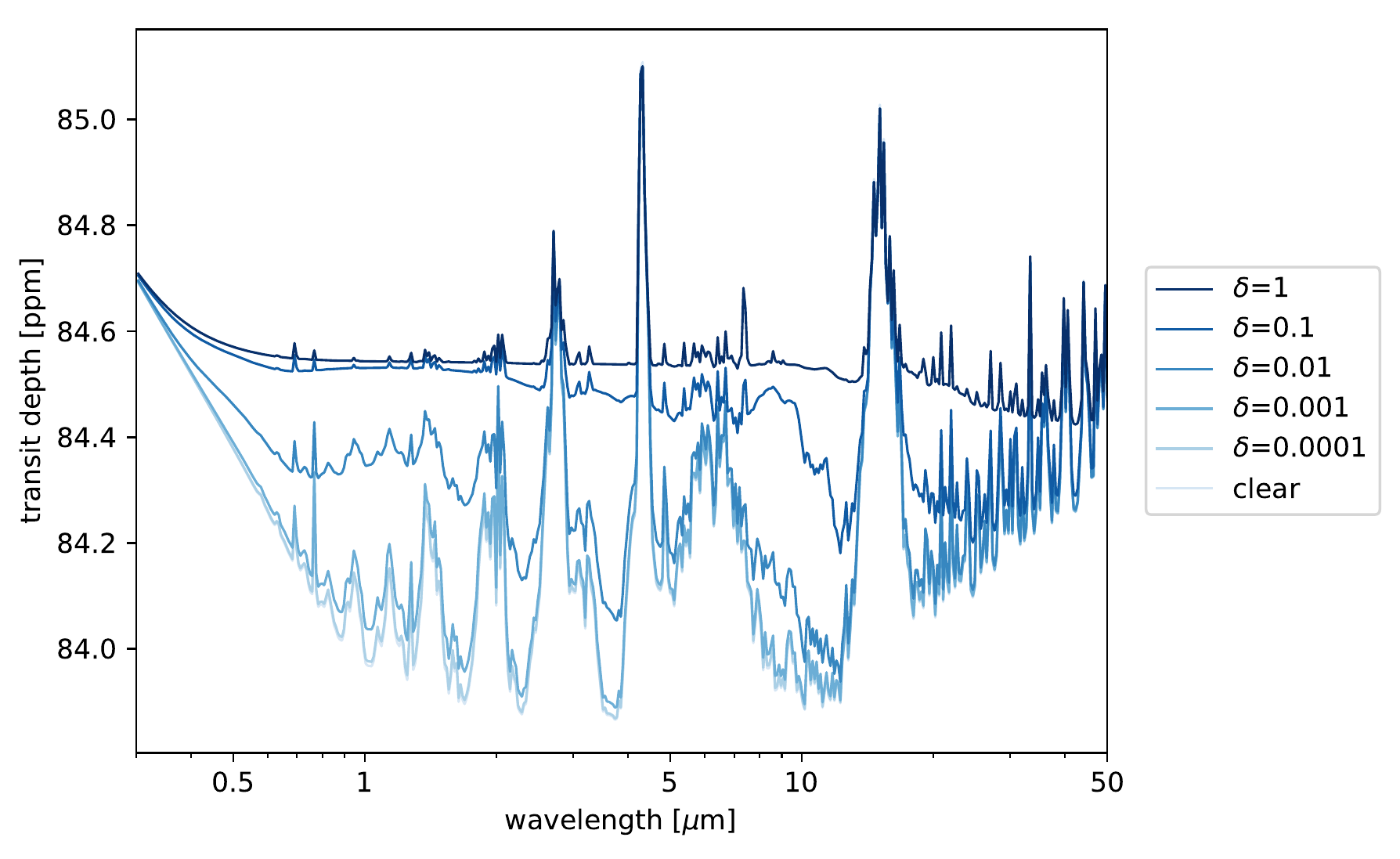}
\caption{Simulated transmission spectra for atmospheres with varying amounts of \ce{H2SO4}-\ce{H2O} aerosols, as measured by their vertical path optical depth $\delta$. The atmospheric composition and other planetary parameters are Earth-like. Size of transit depth signal will vary as relative star-to-planet size varies.}
\label{fig:spectra}
\end{figure*}

We simulated the transmission spectra for an Earth-like planet with \ce{H2SO4}-\ce{H2O} hazes for a large range of vertical optical depths in Figure \ref{fig:spectra}. The slant angle geometry of transit spectroscopy means that the haze layer causes significant differences to the transit spectrum for $\delta$ 
values as low as 0.01 \citep{Fortney2005}.
These results informed our choice of the critical amount of aerosols needed for observable haze detection in Section \ref{sec:methods}. 
Clearly, the presence of the optically thick haze produces a distinct spectrum from a clear terrestrial-type atmosphere. For the Earth-like atmosphere modeled here, the differences are strongest in 3-5 $\mu$m and 7.5-15 $\mu$m bands. In general, which bands have the strongest difference between hazy and clear conditions will vary with atmospheric composition. The height of the signals---and thus their detectability---will also vary strongly with planetary size relative to stellar size and atmospheric scale height. We discuss distinguishing \ce{H2SO4-H2O} from other spectra-flattening agents in Section \ref{subsec:identifying_haze}.

\subsection{Sulfur in the Atmosphere}
We translated our observable sulfur criteria into a critical amount of sulfur in the atmosphere from Equations \eqref{eq:S_atm_crit_so2} and \eqref{eq:S_atm_crit_haze}. For modern-Earth-like planetary conditions and reasonable model parameters given in Table \ref{tab:parameters}, we calculated that observable \ce{SO2} requires $4.9 \times 10^{37}$ S atoms = $2.6 \times 10^{12}$ kg S in the atmosphere (the equivalent of 1 ppm = 1000 ppb \ce{SO2} in an Earth-like atmosphere). Using the limiting model parameters given in Table \ref{tab:parameters} instead gives 1\% of this value ($2.6 \times 10^{10}$ kg S). For Earth-like conditions and reasonable model parameters, a sustained observable haze layer requires $2.5 \times 10^{36}$ S atoms = $1.3 \times 10^{11}$ kg S in the atmosphere (the equivalent of 50 ppb \ce{SO2} in an Earth-like atmosphere). Considering limiting conditions yields 8.5\% of this value for a Sun-like star ($1.1 \times 10^{10}$ kg S) and 13\% for an M dwarf ($1.7 \times 10^{10}$ kg S). For context, Earth's atmosphere currently contains about 10$^8$ kg of sulfur in \ce{SO2} (equivalent to 0.04 ppb \ce{SO2}---if \ce{SO2} were well-mixed) \citep{Brimblecombe1989}, in an environment where anthropogenic emissions are about 10 times natural volcanic sources \citep{Schlesinger2013}. 

\begin{figure*}
\centering
\epsscale{0.8}
\plotone{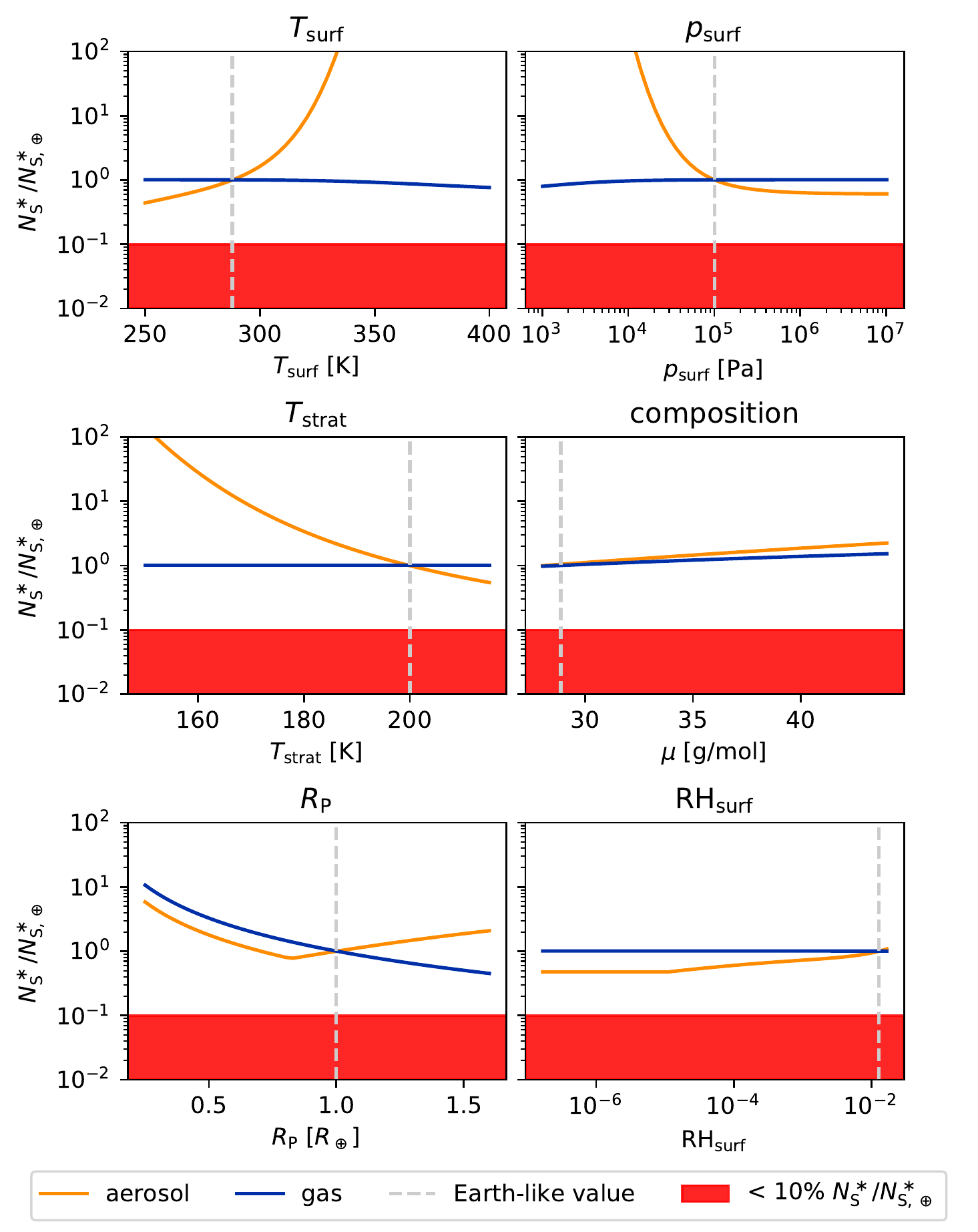}
\caption{Sensitivity of critical number of atmospheric sulfur atoms ($N^\ast_{\ce{S}}$) required for observation for both \ce{SO2} gas (blue) \ce{H2SO4-H2O} aerosols (orange) to planetary conditions of surface pressure ($p_\text{surf}$), surface temperature ($T_\text{surf}$), stratospheric temperature ($T_\text{strat}$), planetary radius ($R_\text{P}$), atmospheric composition, and relative humidity of water vapor at the surface ($\text{RH}_{\text{surf}}$). Modern-Earth-like conditions are marked by a dashed gray vertical line for reference. Solid red boxes indicate regions of high sensitivity ($<$10\% of Earth condition sulfur) to varying planetary parameters. No parameters fall into this regime. Sensitivity to varying atmospheric composition is tested considering only \ce{N2} and \ce{CO2} in the atmosphere and varying relative abundances between them. The resulting average molar mass ($\mu$) is plotted on the x-axis. In testing sensitivity to $R_\text{P}$, we set the mass of the planet $M_\text{P}$ from \cite{Valencia2007}'s scaling relation $R_{\text{P}} \propto M_{\text{P}}^{0.27}$ and include effects from varying surface gravities and outgassing rates.} 
\label{fig:sensitivity}
\end{figure*}

The total number of sulfur atoms required in a planetary atmosphere for observable \ce{SO2} buildup or a haze layer is relatively insensitive to surface temperature $T_\text{surf}$, stratospheric temperature $T_\text{strat}$, surface pressure $p_\text{surf}$, planetary radius $R_\text{P}$, atmospheric composition, and relative humidity of water at the surface $\text{RH}_\text{surf}$ as shown in Figure \ref{fig:sensitivity}. We tested plausible values for these conditions for a terrestrial planet thought to potentially host water:  $T_\text{surf} \in [250 \text{ K}, 400 \text{ K}]$, $T_\text{strat} \in [150 \text{ K}, 225 \text{ K}]$, $p_\text{surf} \in [10^{-2} \text{ atm}, 10^2 \text{ atm}]$, $R_\text{P} \in [0.25 ~R_\oplus, 1.6 ~R_\oplus]$, $\mu \in [28 \text{ g mol}^{-1}, 44 \text{ g mol}^{-1}]$, and $\text{RH}_\text{surf} \in [10^{-5},1]$. Most plausible planetary conditions relative to modern-Earth values actually increase the critical amount of atmospheric sulfur required for observation, which makes our hypothesis of the incompatibility of the observable sulfur stronger. 
Given the lack of sensitivity to planetary conditions, we plot the remaining results assuming Earth-like conditions (i.e., $p_\text{surf}$ = 101325 Pa, $T_\text{surf}$ = 288 K, $T_\text{strat}$ = 200 K, $R_\text{P} = R_\oplus$, composition of air, and $\text{RH}_\text{surf}$ = 0.77).\\

\subsection{Sulfur in the Atmosphere versus Ocean}
\begin{figure}
\centering
\plotone{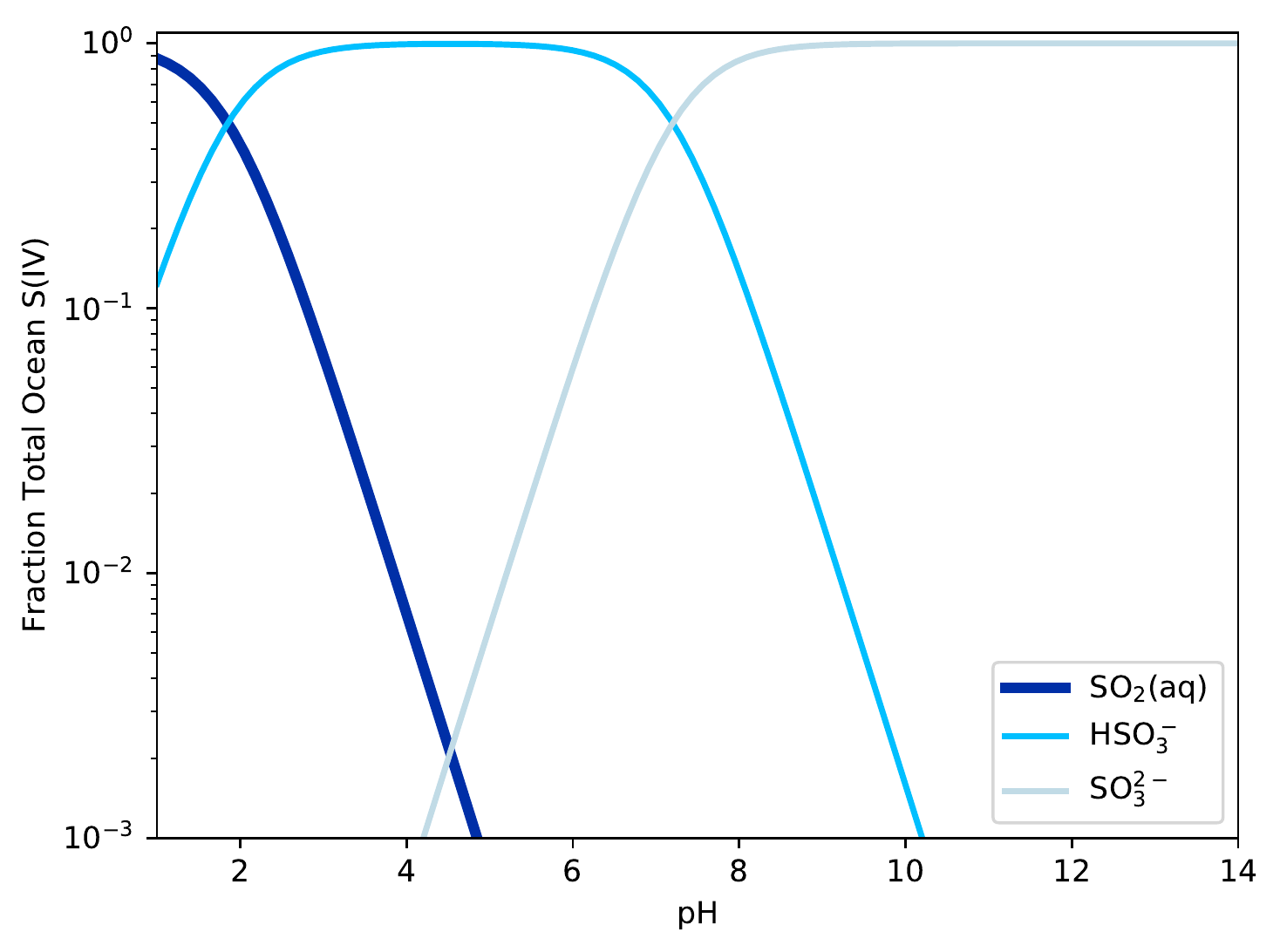}
\caption{Distribution of aqueous S(IV) species \ce{SO2(aq)} (dark blue), \ce{HSO3-} (medium blue), and \ce{SO3^{2-}} (light blue) as a function of pH from the reactions \eqref{eq:aq_react1}-\eqref{eq:aq_react2}. Only \ce{SO2(aq)} is directly in equilibrium with the atmosphere.}
\label{fig:siv_frac}
\end{figure}

We next translated our observable sulfur criteria into a critical amount of sulfur in the ocean-atmosphere system.
We calculated the distribution of sulfur in an ocean between aqueous S(IV) species (\ce{SO2(aq)}, \ce{HSO3-}, \ce{SO3^{2-}}) as a function of pH, assuming S(IV) saturation, in Figure \ref{fig:siv_frac}. The fraction of S(IV) stored as \ce{SO2}---the only dissolved S(IV) species in direct equilibrium with the atmosphere via Equation \eqref{eq:Henry}---exponentially declines as pH increases. For a modern-Earth ocean pH = 8.14, $5.3 \times 10^{-6}$\% of dissolved S(IV) is \ce{SO2(aq)}, 10.3\% is \ce{HSO3-}, and 89.7\% is \ce{SO3^{2-}}. 

\begin{figure}
\centering
\plotone{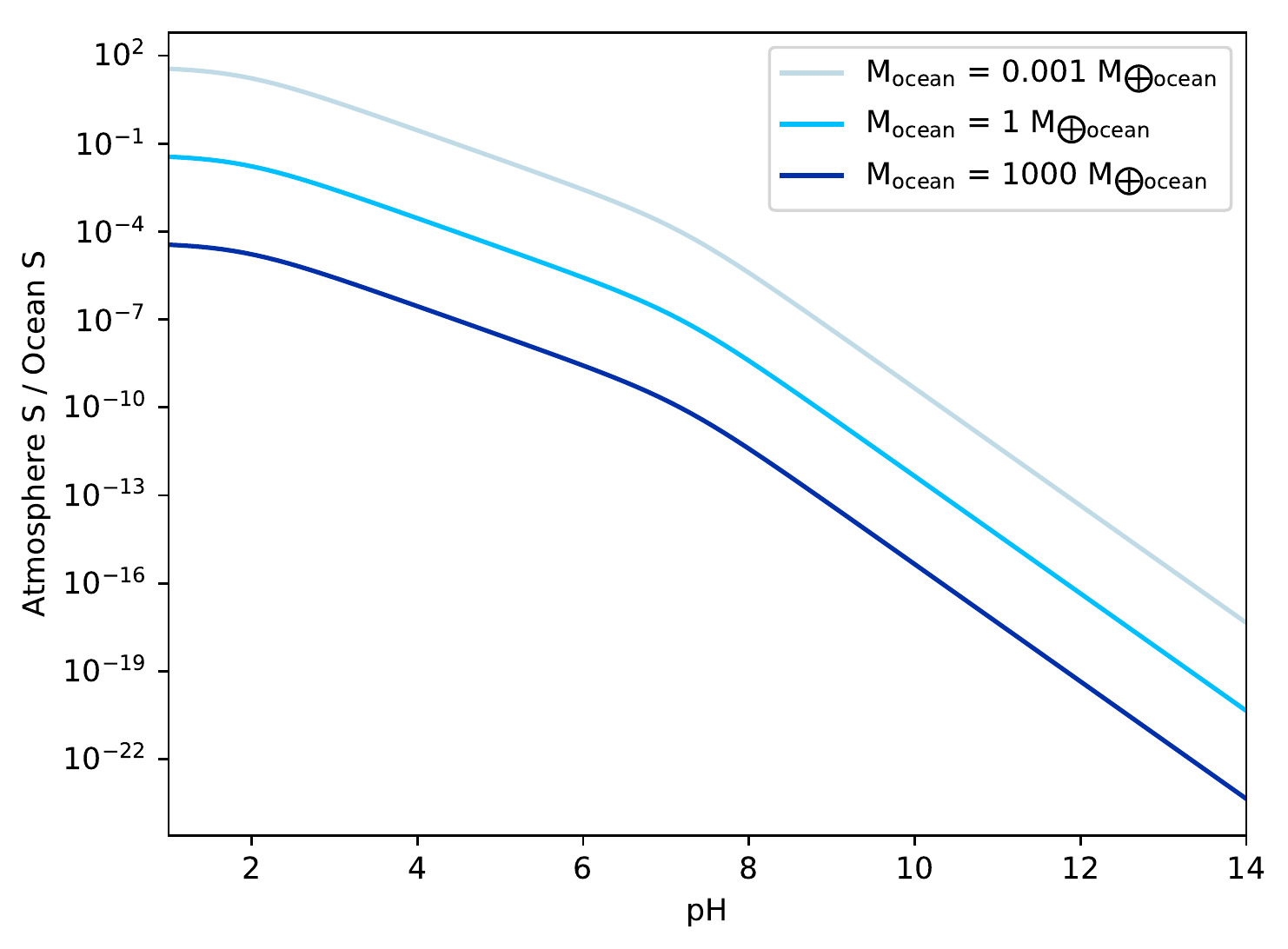}
\caption{The ratio of S in \ce{SO2} in the atmosphere compared to S in S(IV) in the ocean from $N_{\ce{S(IV)}\text{,oc}}$ and $N_{\ce{S}\text{,atm}}$ given in Equations \eqref{eq:S_crit_ocean} and \eqref{eq:S_atm_crit_so2}, respectively, versus ocean pH.}
\label{fig:siv_atm_ocean}
\end{figure}

From the distribution of S(IV) species, we then calculated the expected ratio of S(IV) sulfur in the atmosphere (\ce{SO2}) to S(IV) sulfur in the ocean (\ce{SO2(aq)}, \ce{HSO3-}, \ce{SO3^{2-}}), again assuming saturation of aqueous S(IV) and atmosphere-ocean equilibrium. Figure \ref{fig:siv_atm_ocean} shows the preferential storage of S(IV) in the ocean over the atmosphere as a function of pH with varying total ocean mass. The storage capacity of the ocean linearly increases with increasing ocean mass and exponentially increases with increasing ocean pH.
For Earth ocean pH and mass, the ratio of atmospheric to oceanic S(IV) is $2.13 \times 10^{-9}$. (Note that this ratio is not actually observed in the modern ocean as the assumption of S(IV) saturation is violated because of the instability of aqueous S(IV).)

\subsection{Observable Sulfur versus Ocean Parameters} \label{subsec:obs_sulfur_oc}
Finally, we look at how the presence and characteristics of an ocean shape sulfur observability. First, we calculated conditions for atmospheric sulfur observability for a range of ocean pHs and masses using the best-guess model parameters given in Table \ref{tab:parameters}. Then, we calculated these conditions using the limiting parameters. For all inputs, we plot contours of the critical timescale for S(IV) decay ($\tau_{\ce{S(IV)}}^\ast$) necessary to have enough sulfur in the atmosphere-ocean system to sustain observable \ce{SO2} or an observable \ce{H2SO4-H2O} aerosol layer versus ocean parameters of pH and mass. From present aqueous redox sulfur chemistry experimental results, we set $\tau_{\ce{S(IV)}}^\ast = 0.1$ years $\approx 1$ month (white contour) as a reasonable timescale for aqueous S(IV) decay. For ocean pHs and total masses below this line, observable atmospheric sulfur is possible. Above this line, observable sulfur grows increasingly unlikely. 

Low pH is most favorable for observable atmospheric sulfur in our results, but cations from surface weathering can act to buffer pH and likely maintain a pH near neutral \citep{Grotzinger1993}, particularly in low water content regimes with lots of exposed land. Empirically, paleo-pH estimates for both Earth and Mars seem to favor such a near-neutral regime \citep[e.g.,][]{Halevy2013,Grotzinger2014}. 
From these considerations, we choose pH = 6 to report characteristic ocean sizes leading to observable atmospheric sulfur, given $\tau_{\ce{S(IV)}}^\ast = 0.1$ years. We also choose, somewhat arbitrarily, to define significant surface liquid water as $10^{-3}$ Earth ocean masses or a global equivalent ocean layer of 2.75 m on an Earth-sized planet. For context, on Earth today, this amount of water is equivalent to about one third of the Mediterranean Sea \citep{Eakins2010}.

\begin{figure*}
\centering
\epsscale{0.77}
\plotone{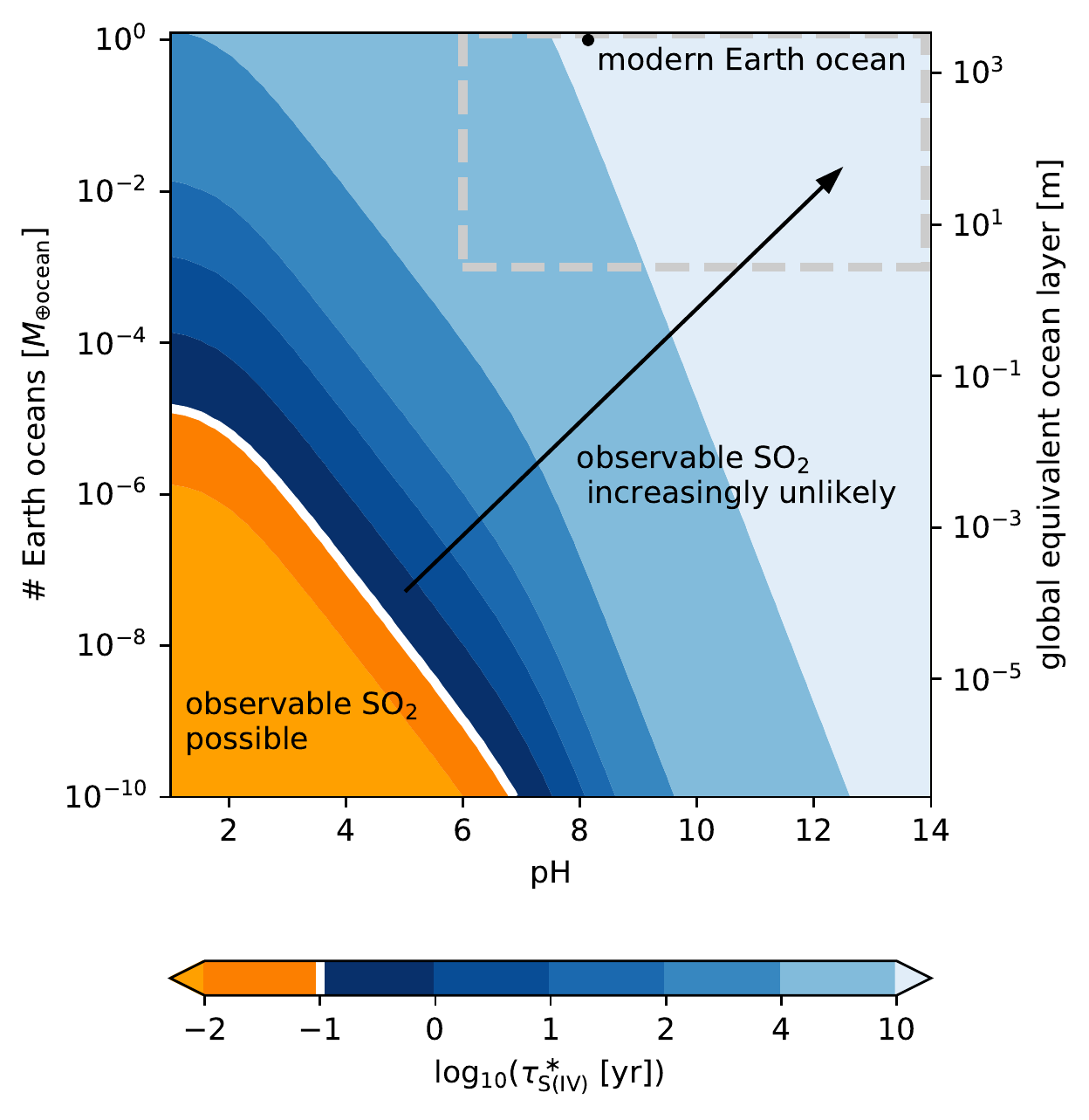}
\caption{Contours of the critical lifetime of aqueous S(IV) (sulfur in redox state +4, in equilibrium with atmosphere) $\tau_{\ce{S(IV)}^\ast}$ for observable mixing ratios of \ce{SO2} versus ocean pH and mass (in Earth ocean masses). Model results are shown for the best-guess model parameters described in the text. The white contour indicates a reasonable timescale for aqueous S(IV) decay. The dashed gray lines indicate ocean parameters of interest (from a reasonable pH limit of 6 and a low-liquid-water threshold of 0.001 $M_{\oplus\text{ocean}}$).}
\label{fig:t_so2_b}
\end{figure*}

Figure \ref{fig:t_so2_b} shows $\tau_{\ce{S(IV)}}^\ast$ versus ocean parameters for observing \ce{SO2} with our best-guess parameter values. For surface liquid water content greater than $10^{-3}$ Earth ocean masses, observable \ce{SO2} with reasonable model parameters requires $\tau_{\ce{S(IV)}}^\ast \gtrsim$ 10 years for even the lowest pHs most favorable for \ce{SO2} buildup and $\tau_{\ce{S(IV)}}^\ast >$ 10,000 years for pH $> 6$. Such long decay times are incompatible with present highest quoted values for decay in the literature, which are on the order of years \citep{Ranjan2018}. At pH = 6 and $\tau_{\ce{S(IV)}}^\ast = 0.1$ years, observable \ce{SO2} requires an ``ocean'' of  less than $1 \times 10^{-9}$ Earth's ocean mass or a global equivalent layer of less than 3 $\mu$m. 

\begin{figure*}
\centering
\epsscale{0.77}
\plotone{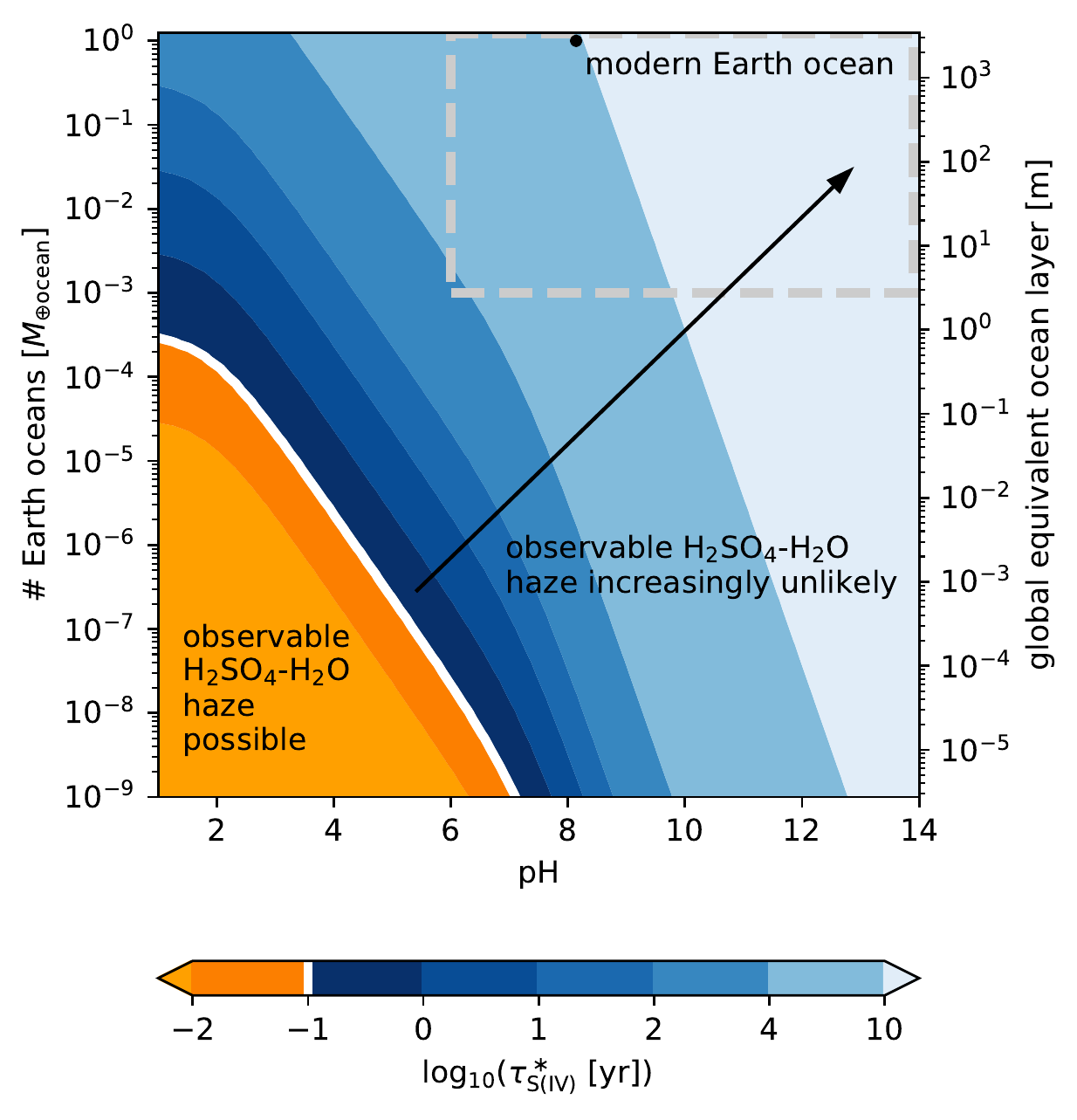}
\caption{Contours of the critical lifetime of aqueous S(IV) $\tau_{\ce{S(IV)}^\ast}$ for the formation of an observable \ce{H2SO4-H2O} haze layer versus ocean pH and mass (in Earth oceans). Model results are shown for the best-guess model parameters described in the text.}
\label{fig:t_b}
\end{figure*}

Figure \ref{fig:t_b} shows $\tau_{\ce{S(IV)}}^\ast$ versus ocean parameters for observing an \ce{H2SO4-H2O} haze layer with our best guess parameter values. For surface liquid water content greater than $10^{-3}$ Earth ocean masses, observable haze formation with reasonable model parameters requires $\tau_{\ce{S(IV)}}^\ast \gtrsim$ 1 year for all pHs and, again, $\tau_{\ce{S(IV)}}^\ast >$10,000 years for pH$> 6$. Again, these timescales are much, much longer than any $\tau_{\ce{S(IV)}}^\ast$ estimates currently present in the literature. At pH = 6 and $\tau_{\ce{S(IV)}}^\ast = 0.1$ years, observable haze requires an ocean of less than $2.5 \times 10^{-8}$ Earth's ocean mass or a global equivalent layer of less than 68 $\mu$m. 

\begin{figure*}
\centering
\epsscale{0.77}
\plotone{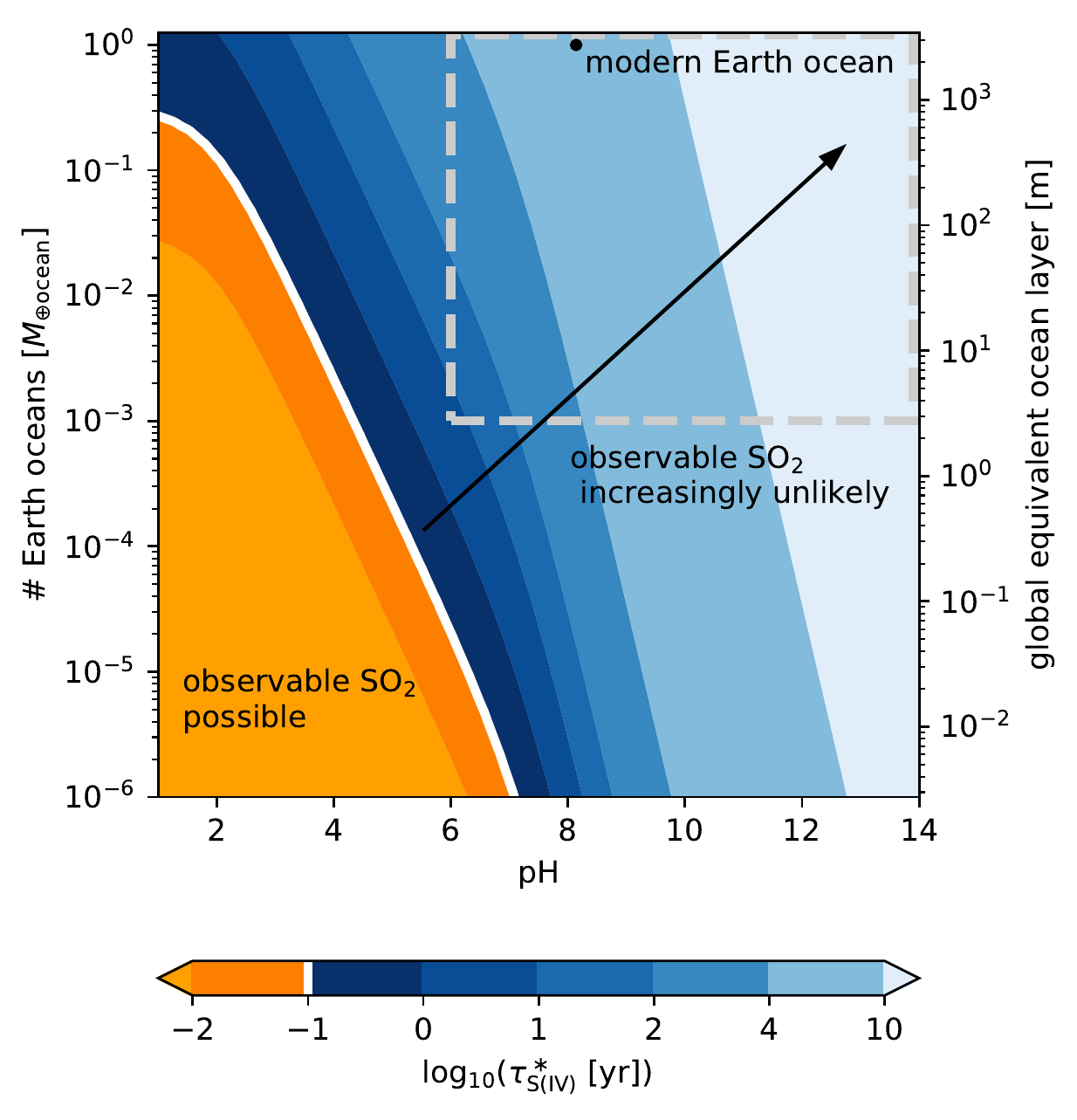}
\caption{Contours of the critical lifetime of aqueous S(IV) $\tau_{\ce{S(IV)}^\ast}$ for the formation of an observable \ce{H2SO4-H2O} haze layer versus ocean pH and mass (in Earth oceans). Model results are shown for the limiting set of model parameters described in the text. }
\label{fig:t_so2_lim}
\end{figure*}

\begin{figure*}
\centering
\plotone{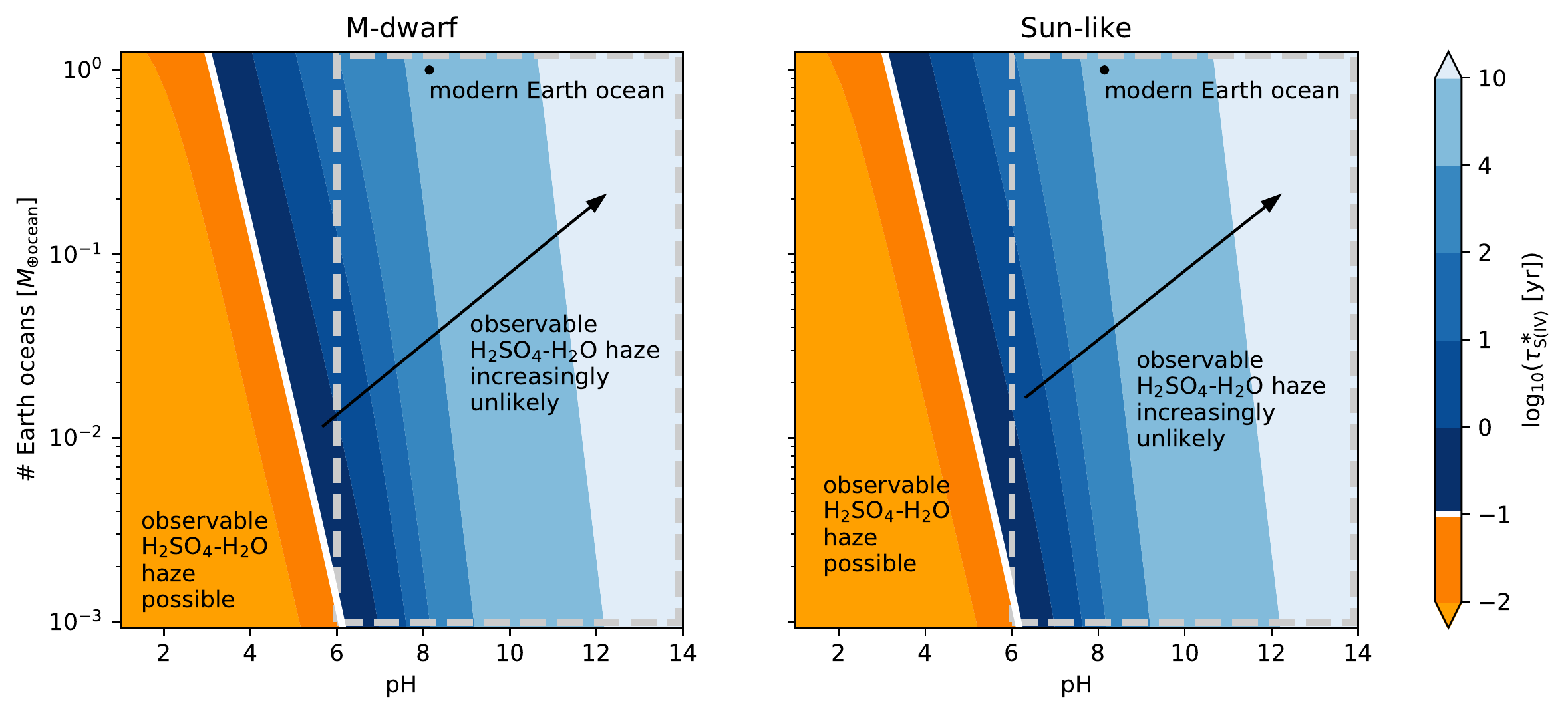}
\caption{Contours of the critical lifetime of aqueous S(IV) $\tau_{\ce{S(IV)}^\ast}$ for the formation of an observable \ce{H2SO4-H2O} haze layer versus ocean pH and mass (in Earth oceans) for a planet around an M dwarf (left) and Sun-like star (right). Model results are shown for the limiting set of model parameters described in the text.}
\label{fig:t_lim}
\end{figure*}

Figures \ref{fig:t_so2_lim} and \ref{fig:t_lim} show $\tau_{\ce{S(IV)}}^\ast$ versus ocean parameters for observing \ce{SO2} and observing a haze layer, respectively, with our limiting parameter values. At pH = 6 and $\tau_{\ce{S(IV)}}^\ast = 0.1$ years, observable \ce{SO2} with these most stringent parameters still requires an ocean of less than $2 \times 10^{-5}$ Earth's ocean mass or a global equivalent layer less than 5.5 cm. We calculate that at pH = 6 and $\tau_{\ce{S(IV)}}^\ast = 0.1$ years, an observable haze layer for a planet around an M-dwarf star requires an ocean of less than $1.3 \times 10^{-3}$ Earth's ocean mass or a global equivalent layer of 3.6 m; around a Sun-like star gives an ocean of less than $1.5 \times 10^{-3}$ Earth's ocean mass or a global equivalent layer of 4.1 m. 

\section{Discussion}\label{sec:dis}
\subsection{Model Uncertainties} 
Robustly establishing sustained high trace mixing ratios of \ce{SO2} and \ce{H2SO4-H2O} haze as indicators of limited surface liquid water will require experimental constraints on S(IV) decay pathways and kinetics. Better theoretical constraints on the upper bounds of sulfur outgassing rates and lower bounds of ocean pHs in water-limited regimes will also improve the robustness of these newly proposed observational constraints on surface liquid water.    

However, in accounting for these poorly constrained inputs, we have concocted a worst-case scenario for each input, often physically inconsistent with each other (e.g., extremely high sulfur outgassing levels would not be consistent with extremely small aerosol particles microphysically). These limiting parameters are intended to push our model to its limits given the breadth of our unknown parameters, rather than just presenting likely values informed by predominantly Earth-based Solar System information. 
The incompatibility of observable \ce{SO2} to below our threshold of significant liquid water by almost two orders of magnitude and the incompatibility of observable \ce{H2SO4-H2O} aerosols to very near our threshold of significant liquid water for these limiting conditions place the extremely strong constraints on surface liquid water for our best-guess model parameters into a larger context.  
Further, many of these model parameters will be more constrainable for individual systems with observations.  Stellar spectrum observations would help constrain the timescale of \ce{SO2} to \ce{H2SO4} conversion $\tau_{\ce{SO2}\rightarrow \ce{H2SO4}}$, and observable stellar properties like rotation rate would help constrain stellar age, which would help place limits on the evolution of outgassing flux $\dot{N_{\ce{S}}}$. 

Though in this paper we have emphasized the benefits of using a simple model for our immediate hypothesis test, in the future a hierarchy of models of varying complexities will be useful. Future work employing more complicated photochemistry, microphysics, aqueous chemistry, and interior dynamics models will allow for more detailed evaluations of the constrainability of surface liquid water from oxidized atmospheric sulfur observations in specific planetary scenarios. Such work would be particularly useful if it were combined with improved experimental constraints on key reaction rates in the atmosphere and ocean.

\subsection{Oxidation State of the Atmosphere}\label{subsec:oxidation_state}
Currently, the proposed atmospheric sulfur anti-ocean signatures are limited in scope to oxidized atmospheres.
Theoretical predictions of the redox evolution of planets orbiting M dwarfs (our most observable targets) suggest such planets are more likely than those orbiting Sun-like stars to evolve toward more oxidized atmospheric conditions, given M-stars' extended high EUV and UV flux period for their first gigayear of main sequence life \citep{Shkolnik2014} and the consequent higher potential for \ce{H2O} dissociation, \ce{H2} escape, and \ce{O2} buildup \citep{Luger2015,Tian2015,Wordsworth2018}. Depending on planetary age and thermal evolution, this \ce{O2} may be simply maintained in the atmosphere, or it may be absorbed into the mantle during a magma ocean phase and influence the oxidation state of the secondary outgassed atmosphere, but either way the planet is driven toward a more oxidized\footnote{We specify that we are referring here to bulk changes in atmospheric redox state as opposed to purely photochemical production of \ce{O2} \citep[e.g.,][]{Tian2014,Domagal2014}, which has been demonstrated to be unlikely given physically motivated planetary conditions \citep{Harman2018b}.} atmosphere \citep{Wordsworth2018}. (Planets with Sun-like host stars can also undergo this process, although in this case its effectiveness is likely more dependent on the composition of the planet's atmosphere \citep{Catling2001,Wordsworth2013,Wordsworth2014}.) 
We can reasonably expect to encounter the oxidized planets that these methods apply to. 
Regardless, the study of the compatibility of buildup of sulfur gas and sulfur aerosol formation with liquid water could be extended to reduced atmospheres (featuring \ce{H2S} gas and \ce{S8} aerosols) after better experimental characterization of the photochemical reactions that produce \ce{S8} and the removal rates of associated dissolved products of \ce{H2S}.

Observationally determining the general oxidation state of a planet's atmosphere will be possible from transit spectra in many cases. \ce{H2}-dominated atmospheres, which are strongly reducing, are identifiable from their extremely high scale heights due to low average molecular mass. \ce{CO2} and \ce{CH4} both have strong spectral features and have the potential to be identified even when not dominant atmospheric gases \citep{Lincowski2018}. While these carbon-bearing species can coexist to some extent, in terrestrial atmospheres, \ce{CO2} is the dominant carbon molecule in oxidized atmospheres and \ce{CH4} or \ce{CO} in reduced atmospheres \citep{Hu2012}. Measuring \ce{CO2}-to-\ce{CH4} (or \ce{CO}) ratios can thus help to constrain an atmosphere's redox state.

\subsection{Identifying Haze as \ce{H2SO4-H2O}}\label{subsec:identifying_haze}
Identifying haze particles as \ce{H2SO4-H2O} aerosols is challenging but tractable. While we leave explicit modeling of haze composition retrieval with stellar noise and specific instrument systematics to future work, we discuss the basic logistics of \ce{H2SO4-H2O} identification here.

\ce{H2SO4}-\ce{H2O} aerosols are only weakly absorbing at visible wavelengths, so Mie scattering dominates their spectral effects. They therefore tend to lack distinctively characteristic spectral fingerprints \citep{Hu2013}. Our simulated transit spectra tests of aerosol cutoff height sensitivity suggest that if the aerosols were to extend throughout much of the otherwise optically thin upper atmosphere ($\gtrsim$ 5 scale heights), \ce{H2SO4-H2O} features could become identifiable. 
Such high extents would seem to be disfavored from \ce{H2SO4} photochemical production and \ce{SO2} vertical transport considerations, but more detailed photochemical modeling and simulated retrievals are required to determine whether this direct identification possibility is actually feasible.
Regardless, \ce{H2SO4}-\ce{H2O} aerosols still have distinct radiative and formation properties when compared to other key spectra-flattening candidates:  water clouds, organic photochemical hazes, and elemental sulfur hazes. 

We tested the ability to distinguish both low-lying water clouds and higher ice clouds (with Earth-like characteristics) from an \ce{H2SO4-H2O} haze layer, but even at our assumed 100\% cloud coverage, the \ce{H2SO4-H2O} haze is clearly distinguishable from \ce{H2O} clouds; the transit spectrum is probing higher optical depths than expected lower cloud peak heights, and even the higher ice clouds are too low to significantly flatten the transit spectrum. The latter two alternative spectra-flattening candidates---\ce{S8} and photochemical haze---require reducing atmospheres to form. Identification of an atmosphere as oxidizing, as described in the preceding section, strongly disfavors these species' formation. Distinguishing \ce{H2SO4}-\ce{H2O} aerosols from \ce{S8} aerosols that form in reduced atmospheres could also be possible from a distinct spectral feature:  \ce{S8} aerosols' radiative properties transition sharply between 0.3 $\mu$m and 0.5 $\mu$m from dominantly attenuating light via absorption to scattering \citep{Hu2013}; \ce{H2SO4-H2O} aerosols do not exhibit this feature. 

Organic photochemical hazes are an active research topic \citep[e.g.,][]{Horst2018}, but many studies suggest that they require \ce{CH4}/\ce{CO2} ratios $\gtrsim$ 0.1 \citep{Dewitt2009}. Spectra indicating significant \ce{CO2} and the absence of substantial mixing ratios of \ce{CH4} would therefore further disfavor organic photochemical haze as the spectra-flattening agent. 
Additionally, organic photochemical hazes form as fractal aggregates \citep{Bar1988} while \ce{H2SO4}-\ce{H2O} aerosols are spherical \citep{Knollenberg1980}. These distinct shapes have different radiative properties \citep{Wolf2010}, which also have the potential to be recovered via inverse modeling \citep{Ackerman2001}. 

\subsection{Life's Impact on the Sulfur Cycle}\label{subsec:life}
We have neglected the influence of life on the sulfur cycle throughout this paper. Of course, constraining liquid water abundance is in the pursuit of finding life beyond Earth, so we would be remiss not to address life's influence on the sulfur cycle as we have presented it here. The most important direct effect would be sulfur-consuming life's capacity to exchange redox states of aqueous sulfur not expected from basic redox reactions \citep{Johnston2011,Kharecha2005}. The thermodynamic instability of aqueous S(IV) species makes them an attractive microbial food source, so even if S(IV) is the waste product of one organism's metabolism, it is likely to be rapidly consumed again and converted back to another redox state. The largest concern for our picture is the potential for metabolic chains to produce dissolved \ce{H2S} species, which will be in equilibrium with atmospheric \ce{H2S} via Henry's law, that could be oxidized to \ce{SO2} in the atmosphere and thus yield a new, unaccounted source of \ce{SO2}. However, this process would serve as an effective recycling term of sulfur that would be a fraction of outgassing. As a less than order of magnitude effect, this recycling is not likely to impact the conclusions of our study, and thus we do not expect these proposed liquid water constraining methods to be invalidated by the presence of life. We also neglect the possibility of intelligent life modifying its environment via sulfur products \citep[e.g.,][]{Caldeira2013}; we leave coupled ecosystem-sulfur cycle studies as an intriguing topic for future investigation. 

\subsection{Extensions to Observational Techniques beyond Transmission Spectroscopy}
Finally, we have focused our efforts in this paper on observations of the sulfur cycle via transmission spectroscopy; however, our analysis could easily be extended to other observational techniques, notably reflected light spectroscopy given our interest in temperate planets. The impact of different observational techniques is in estimating the critical atmospheric sulfur mass that becomes observable. The technique of interest supplies a critical mass path $u^\ast$ or a critical aerosol optical depth $\delta^\ast$ at which \ce{SO2} or \ce{H2SO4-H2O} aerosols, respectively, become observable. Once this calculation is complete, evaluating the feasibility of atmospheric sulfur buildup for a given amount of surface liquid water will be straightforward using our open-source sulfur model.

\section{Conclusion}\label{sec:conclusion}
The presence of liquid water on an oxidized planet strongly influences its sulfur cycle---particularly the planet's ability to sustain an optically thick \ce{H2SO4-H2O} haze layer or a high trace mixing ratio of \ce{SO2} gas. Detectable levels of both \ce{H2SO4-H2O} aerosols and \ce{SO2} gas require \ce{SO2} in the upper atmosphere, but the presence of an ocean restricts the availability of \ce{SO2} in the atmosphere. For expected ocean pHs, exponentially more \ce{SO2} is stored in the ocean than in the atmosphere because of basic chemical properties of aqueous \ce{SO2}. 
Within the ocean, the dissolved products of \ce{SO2} are thermodynamically unstable and thus short-lived. Recent outgassing must supply both the \ce{SO2} present in the atmosphere necessary for observation and the accompanying amount of aqueous sulfur implied by the size of the ocean. 
 
Via a quantitative model of this wet, oxidized sulfur cycle, we have shown that neither observable \ce{H2SO4-H2O} haze layers nor observable levels of \ce{SO2} are likely compatible with significant surface liquid water ($\gtrsim 10^{-3}$ Earth ocean masses). Despite the uncertainties involved in modeling exoplanet processes, this incompatibility seems to persist even in the most extreme physical conditions to promote \ce{SO2} buildup and haze formation. Thus, we propose the observational detection of \ce{H2SO4-H2O} haze and \ce{SO2} gas as two new constraints on surface liquid water.\\
 
\noindent The code for our sulfur cycle model is available at \url{https://github.com/kaitlyn-loftus/alien-sulfur-cycles}. This work was supported by NASA grants 80NSSC18K0829 and NNX16AR86G. KL thanks Matthew Brennan, Junjie Dong, David Johnston, and Itay Halevy for helpful discussions on various aspects of sulfur chemistry. The authors also thank James Kasting for a productive review.

\bibliographystyle{apalike}
\bibliography{bib}{}

\end{document}